\documentclass{jfm}

\usepackage[dvipsnames]{xcolor}
\usepackage{graphicx}
\usepackage{newtxtext}
\usepackage{newtxmath}
\usepackage{natbib}
\usepackage{hyperref}
\hypersetup{
    colorlinks = true,
    urlcolor   = blue,
    citecolor  = black,
}

\newcommand{\RomanNumeralCaps}[1]

\usepackage{subfig}
\usepackage{amsmath}
\usepackage{epstopdf}
\usepackage{multicol}
\usepackage{amssymb}
\usepackage{textcomp}
\usepackage{color}

\def\rmd{{\rm d}}
\def\rme{{\rm e}}
\def\rmi{{\rm i}}
\def\rmj{{\rm j}}

\def\diff#1#2{{{\rmd #1} \over {\rmd#2}}}

\newcommand{\markblue}[1]{\textcolor{black}{#1}}

\title[]{Vertical motion of a periodically driven floating disc}
\author{Anand U. Oza\aff{1}\corresp{\email{oza@njit.edu}}, Jack-William Barotta\aff{2,3}, Eli Silver\aff{2} and Daniel M. Harris\aff{2}\corresp{\email{daniel\_harris3@brown.edu}}}

\affiliation{
\aff{1}Department of Mathematics \& Center for Applied Mathematics and Statistics, New Jersey Institute of Technology, Newark, New Jersey 07102, USA
\aff{2}School of Engineering, Center for Fluid Mechanics, Brown University,
184 Hope Street, Providence, Rhode Island 02912, USA
\aff{3}Franklin W. Olin College of Engineering, 1000 Olin Way, Needham, MA 02492, USA}

\begin{document}
\maketitle

\abstract{
We present the results of a combined theoretical and experimental investigation into the vertical dynamics of floating discs subjected to an imposed time-periodic forcing. The axisymmetric and inviscid wavefield is governed by a linear elliptic boundary value problem with mixed boundary conditions, wherein the no-penetration boundary condition is satisfied under the disc while the free surface boundary conditions are enforced away from it. The problem is solved by recasting the system of partial differential equations as a second-kind Fredholm integral equation which is then solved numerically. The solution furnishes a prediction for the dependence of the disc's oscillation amplitude on the forcing frequency, which exhibits \markblue{good} agreement with experiments. We interpret our results physically by computing the added mass, wave damping and effective spring coefficients of the disc, both numerically for a range of forcing frequencies and analytically in the low-frequency limit.
}

\section{Introduction}\label{Sec:Intro}

Self-propulsion is the remarkable property of some entities that use energy from the environment or internally generated to set themselves into persistent motion. Over the last two decades, attention has been given to the self-propulsion of bodies, both living~\citep{bush2006walking} and non-living~\citep{Yuan2012}, at the liquid-gas interface. More recently, an experimental study on so-called {\it capillary surfers} demonstrated that millimetric hydrophobic bodies could self-propel while floating on the surface of a vertically vibrating fluid bath~\citep{ho2023capillary}. In that experiment, the fluid bath is driven well below the Faraday instability threshold~\citep{faraday1831forms}, so the vertical oscillations of the body are responsible for generating gravity-capillary waves at the interface. The body's shape is fore-aft asymmetric, which leads to an asymmetric interfacial wavefield and associated fluid flow. 
The so-called ``SurferBot" self-propels by a similar mechanism: it is pinned at the air–water interface and driven by a vibration motor that is positioned so as to generate an asymmetric rocking motion, which causes it to self-propel steadily at speeds on the order of centimeters per second~\citep{Rhee2022}. The design of the SurferBot was inspired by the honeybee, which, if trapped at an air-water interface, can rapidly oscillate its wings to propel itself forward in an attempt to reach safety~\citep{roh2019honeybees}. 
Subsequent studies on {\it capillary spinners}~\citep{Barotta2023,Barotta2024,Sungar2024} demonstrated that the waves generated by a floating chiral object can lead to its geometrically induced steady rotation about its center of mass. Taken together, these studies demonstrate that periodic heaving motion can lead to persistent horizontal motion or rotation at the interface, with novel collective effects enabled via wave-mediated stresses at the interface~\citep{Harris2025Review}.

The basic mechanism by which a capillary surfer self-propels is that it generates an asymmetric wavefield, which leads to a momentum flux and thus a net propulsive force on the surfer \citep{longuet1964radiation, longuet1977mean}. However, there does not exist a systematic analytical theory that describes the waves and flows generated by asymmetric objects that oscillate while floating at the interface. A recent numerical study sought to characterize the waves generated by a tilted rigid plate floating on the surface of a two-dimensional (2D) fluid bath~\citep{Benham2024}. The plate was subjected to time-periodic forces in the horizontal and vertical directions, and the hydrodynamic force on the plate was computed. In that model, the wavefield was described by the so-called quasipotential model, a theory valid in the regime of weak viscosity wherein the waves are assumed to be irrotational and inviscid at leading order, but dissipation in the boundary layer at the free surface is accounted for~\citep{Lamb1932,Dias2008}. 

As a first step towards understanding the interfacial self-propulsion of surfers, we consider here the simpler scenario of a symmetric object, which necessarily will not move horizontally but will still generate waves and flows if it oscillates vertically. We will make two simplifications for the sake of tractability: first, we will neglect viscosity, an assumption that we expect to be valid if the characteristic size of the boundary layer $\sqrt{\nu/\omega}$ is much less than the wavelength $\lambda$ \markblue{and disc radius $R$}, where $\nu$ is the kinematic viscosity of the fluid and $\omega$ is the angular frequency of oscillation. Second, we will linearize the equations of motion, an assumption that is valid provided that the amplitude of oscillation is small relative to the wavelength. Of particular interest are the added (or virtual) mass and wave damping (or resistance) coefficients of the body, and the dependence of both on the forcing frequency. Intuitively, the added mass is associated with the fluid displaced by the body as it oscillates, and the damping coefficient indicates the degree to which the body effectively loses energy by radiating waves. Note that radiation is the only source of damping in this problem because viscosity is neglected. 

There is an extensive theoretical and numerical literature on the linear gravity waves generated by a floating body~\citep{Wehausen1971}, which is applicable if surface tension can be neglected, that is, if the wavelength and the body's characteristic size $R$ are large relative to the capillary length $\sqrt{\sigma/(\rho g)}$, where $\sigma$ is the surface tension, $\rho$ the fluid density and $g$ the gravitational acceleration. The associated mathematical problem is to find a scalar velocity potential $\phi$ that satisfies the no-penetration boundary condition on the body, the dynamic and kinematic boundary conditions on the free surface, and a radiation condition at infinity. The gravity wavefields generated by a half-immersed, periodically-heaving horizontal cylinder in 2D~\citep{Ursell1949} and sphere in 3D~\citep{Havelock1955,Hulme1982} have been studied theoretically by {\markblue{expressing the solution as a multipole expansion}}. 
Doing so yields an infinite system of equations for the coefficients of that series which can be truncated and solved numerically. The potential flow generated by a heaving disc was first computed by~\citet{MacCamy1961a}, who derived an integral equation and then solved it numerically. A similar method was used to find the added mass and damping coefficients of a pitching disc~\citep{Kim1963}, which was subsequently generalized to treat an immersed ellipsoid executing arbitrary time-periodic oscillations in 3D~\citep{Kim1965}. \citet{Miles1987} also studied the heaving motion of a floating disc by recasting the problem using the Hankel transform, and using a variational method to approximately solve the resulting integral equation. The method of matched eigenfunction expansions was used to find the added mass and damping coefficients of a heaving vertical cylinder partially submerged in a bath of finite depth~\citep{yeung1981added}. {\markblue{There is also a body of work on the scattering of waves by fully submerged thin discs, which we review only partially here. The problem has been solved by matched eigenfunction expansions, for both a rigid~\citep{Yu1993} and porous~\citep{Chwang1994} disc; by recasting the problem as a hypersingular integral equation~\citep{Martin1997}; and through use of the Hankel transform~\citep{Porter2015}. Two-layer fluids have also been considered, with prior work on the scattering of waves by a rigid circular disc~\citep{Islam2019}, rigid elliptical disc~\citep{Das2023} and porous circular disc~\citep{Das2025}.}}

The influence of surface tension on the waves and flows generated by oscillating bodies has received comparatively less attention. Theories have been developed for the 2D gravity-capillary waves generated by a half-immersed heaving horizontal cylinder~\citep{Evans1968b} and a ``wavemaker," or a laterally-oscillating vertical plate~\citep{RhodesRobinson1971,Mandal1997}. In those theories, the slope of the free surface at the contact line is prescribed. Models that incorporate contact-angle hysteresis have also been developed, wherein the velocity of the free surface at the contact line relative to the body is assumed to be proportional to the slope. Two special cases of such models are that of a freely moving contact line with a contact angle fixed at 90$^\circ$, and a fixed contact line, as assumed herein. For example, the waves generated by a plate moving periodically into and out of a liquid bath~\citep{Young1987}, a heaving cylinder of arbitrary cross-section~\citep{Hocking1988a}, and wavemaker~\citep{Hocking1991,Yeh2008} have been studied theoretically. While the aforementioned studies are restricted to 2D, the related problem of gravity-capillary wave scattering off a submerged barrier has been studied in 3D, specifically, for waves impinging obliquely on a wall~\citep{Hocking1988b}, and for waves scattering off an infinitely deep vertical cylinder~\citep{Hocking1990}. 
More recently,~\citet{OzaSurfers} used the quasipotential model~\citep{Lamb1932,Dias2008} to find an analytical solution for the weakly viscous gravity-capillary wavefield generated by an oscillating point source at the interface, thus generalizing prior work that considered the capillary wavefield generated by a point source on the surface of an inviscid bath~\citep{de2018capillary}. \citet{OzaSurfers} used their point-source solution to construct a theoretical model that represented a surfer as a pair of point sources of interfacial gravity-capillary waves, which allowed them to rationalize the bound states adopted by pairs and larger collectives of surfers in experiments~\citep{ho2023capillary}.

We here investigate the vertical dynamics of a periodically driven floating disc through a combination of theory and experiment. The disc is subjected to a prescribed oscillatory force in the vertical direction, which leads to its heaving motion. The amplitude and phase response of that heaving motion is determined by the balance between the disc's inertia, its weight, the applied force, and the hydrodynamic forces associated with buoyancy, dynamic pressure and surface tension. 
The dynamic pressure is due to the gravity-capillary waves radiated by the disc as it oscillates. These waves also contribute to the surface tension force, which has a static contribution due to the pinned contact line.
The theoretical problem is thus to find the gravity-capillary wavefield generated by a heaving disc. {\markblue{For the sake of simplicity, we assume the fluid to be inviscid and of infinite depth, and we linearize the governing equations.}} 
Notably, the radius of the disc is assumed to be comparable to the capillary length, which requires considering both surface tension and gravitational forces. Our work thus generalizes prior work in which the added mass and wave damping coefficients of a floating disc were computed in the {\markblue{absence of surface tension}}~
\citep{MacCamy1961a,Miles1987}. In the experiment, a magnetic field is used to generate sinusoidal forcing directly on the floating disc and the resulting steady-state kinematic response is measured directly. We investigate the dependence of the disc's oscillation amplitude and phase response on the forcing frequency, and compare our measurements to the predictions of the theory.

The paper is organized as follows. The theory is described in \S~\ref{Sec:model}: the model is given in \S~\ref{SSec:model}, which is reduced to an integral equation in \S~\ref{SSec:IntEq} through use of the Hankel transform. The method for solving the integral equation is described in \S~\ref{SSec:Numerical}, and a simplification for low forcing frequencies is presented in \S~\ref{SSec:LowFreq}. The experimental protocol is detailed in \S~\ref{Sec:Exp}. The comparison between theory and experiment is given in \S~\ref{Sec:ExpTheory}. A physical interpretation of the results is given in \S~\ref{Sec:Physical}, wherein the added mass and resistance coefficients are computed and compared to their gravity-wave counterparts~\citep{MacCamy1961a,Miles1987}.

\section{Theoretical modeling}\label{Sec:model}

\subsection{Governing equations}\label{SSec:model}

\begin{figure}
\begin{center}
\includegraphics[width=.85\textwidth]{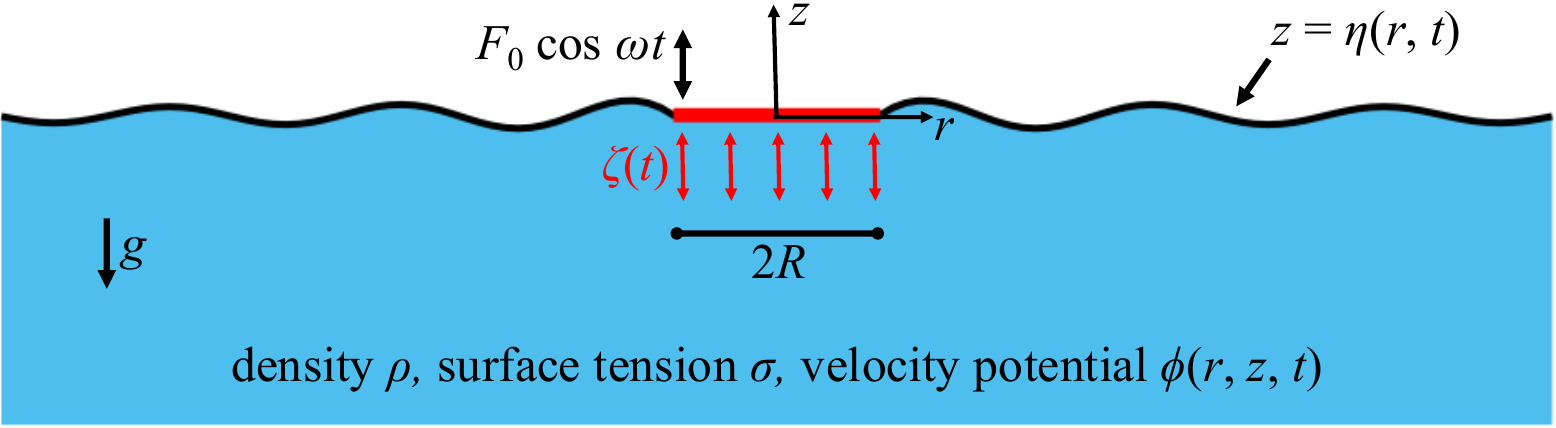}
\caption{Schematic of the mathematical problem solved in \S~\ref{Sec:model}. \markblue{A disc (red) of radius $R$ floats on the surface of an infinitely deep fluid bath (blue) of density $\rho$ and surface tension $\sigma$ in the presence of a gravitational acceleration $g$. The disc is subjected to an oscillatory force of amplitude $F_0$ and frequency $\omega$ in the vertical direction. The vertical position of the disc is $\zeta(t)$, the velocity potential of the fluid is $\phi(r,z,t)$, and the height of the wavefield is given by $z=\eta(r,t)$.}\label{fig:schematic}
}
\end{center} 
\end{figure}

Consider a solid disc of mass $m$, radius $R$ 
and vertical position $\zeta(t)$ floating on the surface of an inviscid, incompressible, irrotational and infinitely deep fluid bath of density $\rho$ and surface tension $\sigma$ in the presence of a gravitational acceleration $g$ (figure~\ref{fig:schematic}). The disc is subjected to a time-periodic vertical force of amplitude $F_0$ and angular frequency $\omega$. The disc's weight $mg$ and applied force magnitude $F_0$ are assumed to be small relative to the maximum capillary force $2\pi R\sigma$, approximations that allow us to linearize the governing equations around the flat interface, $z=0$. The disc generates a small-amplitude (linear) axisymmetric gravity-capillary wavefield $\eta(r,t)$ and a fluid velocity field described by the velocity potential $\phi(r,z,t)$, where $r=\sqrt{x^2+y^2}$ and $z\leq 0$. The governing equations for the fluid are
\begin{subequations}
\begin{alignat}{2}
\Delta\phi+\partial_{zz}\phi&=0, &&\markblue{r\geq 0,\quad z < 0}, 
\label{Laplace}\\
\partial_t\phi&=-g\eta+\frac{\sigma}{\rho}\Delta \eta,\quad &&\markblue{r > R,\quad z=0},\label{DynBC}\\
\partial_t\eta&=\partial_z\phi,\quad &&\markblue{r > R,\quad z=0}, \label{KinBC}\\
\partial_z\phi&=\dot{\zeta},&&\markblue{r < R,\quad z=0}, \label{NoPenBC}\\
\eta(R,t)&=\zeta,\label{PinBC} \\
\phi&\rightarrow 0\quad&&\text{as}\quad z\rightarrow -\infty,\label{zInfBC} \\
\eta&\rightarrow 0\quad &&\text{as}\quad r\rightarrow\infty,\label{rInfBC}
\end{alignat}\label{BathEqn}
\end{subequations} 
where $\Delta=r^{-1}\partial_r(r\partial_r)$ is the radial Laplacian. 
Equation~\eqref{Laplace} enforces incompressibility of the fluid, and Eqs.~\eqref{DynBC} and~\eqref{KinBC} are the dynamic and kinematic boundary conditions at the free surface, respectively. Equation~\eqref{NoPenBC} is the no-penetration boundary condition on the underside of the disc. Equation~\eqref{PinBC} requires that the fluid interface remains pinned to the disc at the contact line, an assumption that is consistent with the experiments and physically reasonable because the disc is made of a hydrophobic material. The disc moves in response to the gravitational force, buoyancy, surface tension force, dynamic pressure and applied force, so its vertical dynamics is given by
\begin{align}
m\ddot{\zeta}=-mg-\rho g\pi R^2\zeta+ 2\pi \sigma R\partial_r\eta(R,t)-2\pi\rho\int_0^R\rmd r\,r\partial_t\phi|_{z=0}+F_0\cos\omega t.\label{ZEqn}
\end{align}

We non-dimensionalize Eqs.~\eqref{BathEqn}-\eqref{ZEqn} using the lengthscale $R$ and timescale $\sqrt{R/g}$, which introduces the dimensionless parameters
\begin{align}
\Omega = \omega\sqrt{\frac{R}{g}},\quad \text{Bo} = \frac{\rho g R^2}{\sigma}\quad\text{and}\quad M=\frac{m}{\rho R^3},\label{NDimParam}
\end{align}
which are the dimensionless forcing frequency, Bond number and dimensionless mass of the disc, respectively. We seek time-harmonic solutions to the resulting dimensionless equations of the form 
\begin{align}
\phi(r,z,t)=\text{Re}\left[\Phi(r,z)\rme^{\rmi\Omega t}\right],\quad \eta(r,t)=\text{Re}\left[H(r)\rme^{\rmi\Omega t}\right],\quad \zeta(t)=\text{Re}\left[Z\rme^{\rmi\Omega t}\right]
\end{align}
and thus obtain the following linear elliptic boundary value problem for the bath:
\begin{subequations}
\begin{alignat}{2}
\Delta\Phi+\partial_{zz}\Phi&=0,\quad &&r \geq 0,\quad z < 0,\label{LaplaceND}\\
\mathrm{i}\Omega\text{ Bo }\Phi&=-\text{Bo }H+\Delta H,\quad &&r > 1,\quad z =0,\label{DynBCND}\\
\mathrm{i}\Omega H&=\partial_z\Phi,\quad &&r > 1,\quad z = 0,\label{KinBCND}\\
\partial_z\Phi&=\mathrm{i}\Omega Z,\quad &&r < 1,\quad z =0,\label{NoPenBCND} \\
H(1)&=Z,\label{PinBCND} \\
\Phi&\rightarrow 0\quad&&\text{as}\quad z\rightarrow-\infty,\label{zInfBCND} \\
\sqrt{r}\left(\partial_r H+\rmi q_{\text{r}}H\right)&\rightarrow 0\quad&&\text{as}\quad r\rightarrow\infty,\label{RadCond}
\end{alignat}\label{BathEqnND}
\end{subequations}
where $q_{\text{r}}\in\mathbb{R}$ and $q_{\text{i}}\in\mathbb{C}$ are the roots of the deep-water gravity-capillary dispersion relation
\begin{align}
q(q^2+\text{Bo})-\text{Bo }\Omega^2=(q-q_{\text{r}})(q-q_{\text{i}})(q-\overline{q}_{\text{i}}).\label{DispRel} 
\end{align}
Equation~\eqref{RadCond} is a radiation condition at infinity. {\markblue{We note that while the time-domain decay condition~\eqref{rInfBC} would map to $H(r) \to 0$, this simple decay is insufficient to guarantee a unique solution in the steady-state frequency domain, as both incoming and outgoing cylindrical waves decay as $O(r^{-1/2})$. The radiation condition enforces causality, isolating the outgoing wave solution to ensure that waves generated by the disc propagate strictly outwards.} 

The disc's (complex) oscillation amplitude follows from~\eqref{ZEqn}:
\begin{align}
Z
&=\frac{F_0/(\rho gR^3)}{-(M+M_{\text{P}})\Omega^2+\rmi\Omega\left(C_{\text{ST}}+C_{\text{P}}\right)+k_{\text{ST}}+\pi},\label{ZEqnND}
\end{align}
where we define
\begin{align}
-2\pi\rmi\int_0^1\rmd r\,r\Phi(r,0)/Z=\Omega M_{\text{P}}-\rmi C_{\text{P}},\quad -2\pi\text{ Bo}^{-1}H^{\prime}(1)/Z=k_{\text{ST}}+\rmi\Omega C_{\text{ST}}.\label{CoeffDef}
\end{align}
The physical interpretation of these terms is that the disc acquires an added mass $M_{\text{P}}$ due to the dynamic pressure, and that surface tension at the contact line exerts an effective linear spring force on the disc with spring constant $k_{\text{ST}}$. Additionally, both pressure and surface tension forces dampen the disc's oscillations with damping coefficients $C_{\text{P}}$ and $C_{\text{ST}}$, respectively. These damping terms arise because the disc effectively loses energy to the waves that it radiates as it oscillates. As a reminder, the energy loss due to radiation is the only source of damping in this model because viscosity is neglected. Note that the quantities $M_{\text{P}}$ and $C_{\text{P}}/\Omega$ associated with the dynamic pressure were called $\hat{M}$ and $\hat{R}$, respectively, by~\cite{Miles1987} in his study of the gravity waves generated by an oscillating disc. 

\subsection{Reduction to an integral equation}\label{SSec:IntEq}

We proceed by finding an expression for the wave height $H$ in terms of the velocity potential $\Phi$. Equation~\eqref{DynBCND} is a screened Poisson equation for $H$, whose Green's function $\mathcal{G}(r,\xi)$ is
\begin{align}
\mathcal{G}(r,\xi) &= \frac{\xi}{\text{K}_0\left(\sqrt{\text{Bo}}\right)}\begin{cases} 
G(r)\text{K}_0\left(\xi\sqrt{\text{Bo}}\right) &\text{if }r < \xi, \\ G(\xi)\text{K}_0\left(r\sqrt{\text{Bo}}\right) &\text{if }r > \xi, 
\end{cases}
\nonumber \\
\text{where}\quad G(r)&=\text{K}_0\left(r\sqrt{\text{Bo}}\right)\text{I}_0\left(\sqrt{\text{Bo}}\right)-\text{I}_0\left(r\sqrt{\text{Bo}}\right)\text{K}_0\left(\sqrt{\text{Bo}}\right)\nonumber
\end{align}
and $\text{I}_n(r)$ and $\text{K}_n(r)$ are the modified Bessel functions of the first and second kind of order $n$, respectively. To verify that $\mathcal{G}(r,\xi)$ is the Green's function, note first that $G(1)=0$, as required by~\eqref{PinBCND}, and that $\text{K}_0$ decays at infinity. To evaluate the jump in $\partial\mathcal{G}/\partial r$ at $r=\xi$, recall that $\text{K}_0^{\prime}(r)=-\text{K}_1(r)$, $\text{I}_0^{\prime}(r)=\text{I}_1(r)$ and
\begin{align}
\text{K}_0(r)\text{I}_1(r)+\text{I}_0(r)\text{K}_1(r)&=\frac{1}{r},\label{Fact3}
\end{align}
which together imply that
\begin{align}
\sqrt{\text{Bo}}\,G(r)\text{K}_1\left(r\sqrt{\text{Bo}}\right)+G^{\prime}(r)\text{K}_0\left(r\sqrt{\text{Bo}}\right)=-\frac{\text{K}_0\left(\sqrt{\text{Bo}}\right)}{r},\label{Fact4}
\end{align}
and thus that $\partial\mathcal{G}/\partial r(\xi^+,\xi)-\partial\mathcal{G}/\partial r(\xi^-,\xi)=1$. Having verified that $\mathcal{G}(r,\xi)$ is the Green's function of Eq.~\eqref{DynBCND}, the solution for $H(r)$ follows directly:
\begin{align}
H(r)&=\frac{\rmi\Omega\text{ Bo}}{\text{K}_0\left(\sqrt{\text{Bo}}\right)}\left[\text{K}_0\left(r\sqrt{\text{Bo}}\right)\int_1^r\rmd\xi\,\xi\Phi(\xi,0)G(\xi)+G(r)\int_r^{\infty}\rmd\xi\,\xi\Phi(\xi,0)\text{K}_0\left(\xi\sqrt{\text{Bo}}\right)\right]\nonumber \\
&\phantom{=}+Z\frac{\text{K}_0\left(r\sqrt{\text{Bo}}\right)}{\text{K}_0\left(\sqrt{\text{Bo}}\right)}.\label{h1Int1}
\end{align}

We now introduce the Hankel transform in $r$ of the velocity potential $\Phi$. Equation~\eqref{LaplaceND}, combined with the decay condition~\eqref{zInfBCND}, 
may readily be solved to yield
\begin{align}
\Phi(r,z)=\int_0^{\infty}\mathrm{d}k\,kA(k)\text{J}_0(kr)\mathrm{e}^{kz},\label{PhiHankel}
\end{align}
where $A(k)$ is unknown. Substituting~\eqref{PhiHankel} into~\eqref{h1Int1}, interchanging the order of integration and using the facts that
\begin{subequations}
\begin{alignat}{1}
\int_1^r\rmd\xi\,\xi\text{J}_0(k\xi)G(\xi)
&=\frac{r}{\text{Bo}+k^2}\left[\frac{\text{J}_0(k)}{r}+\text{J}_0(kr)G^{\prime}(r)+k\text{J}_1(kr)G(r)\right],\label{Fact1}\\
\int_r^{\infty}\rmd\xi\,\xi\text{J}_0(k\xi)\text{K}_0\left(\xi\sqrt{\text{Bo}}\right)&=\frac{r}{\text{Bo}+k^2}\left[\sqrt{\text{Bo}}\,\text{K}_1\left(r\sqrt{\text{Bo}}\right)\text{J}_0(kr)-k\text{K}_0\left(r\sqrt{\text{Bo}}\right)\text{J}_1(kr)\right], \label{Fact2}
\end{alignat}\label{Facts}
\end{subequations}
Eq.~\eqref{h1Int1} reduces to
\begin{align}
H(r) = -\mathrm{i}\Omega\text{ Bo}\int_0^{\infty}\mathrm{d} k\,\frac{k}{\text{Bo}+k^2}A(k)\left[\text{J}_0(kr)-\frac{\text{K}_0\left(r\sqrt{\text{Bo}}\right)}{\text{K}_0\left(\sqrt{\text{Bo}}\right)}\text{J}_0(k)\right]+Z\frac{\text{K}_0\left(r\sqrt{\text{Bo}}\right)}{\text{K}_0\left(\sqrt{\text{Bo}}\right)},\label{h1Soln}
\end{align}
where we have used~\eqref{Fact4} to simplify the integrand.

To find $A(k)$, we combine the kinematic boundary condition at the free surface~\eqref{KinBCND} and the no-penetration boundary condition on the disc~\eqref{NoPenBCND}, to yield
\begin{align}
\int_0^{\infty}\mathrm{d}k\,A(k)k^2\text{J}_0(kr)=\begin{cases} \mathrm{i}\Omega Z &\text{if }r < 1, \\ \mathrm{i}\Omega H(r)&\text{if }r > 1.\end{cases}\label{BCComb1}
\end{align}
We multiply both sides of Eq.~\eqref{BCComb1} by $r\text{J}_0(qr)$ and integrate on $0\leq r < \infty$. Interchanging the order of integration and using the fact that
\begin{align}
\int_0^\infty\rmd r\, r\text{J}_0(kr)\text{J}_0(qr)&=\frac{\delta(k-q)}{k},
\end{align}
the left-hand side of~\eqref{BCComb1} reduces to $A(q)q$. Using~\eqref{Fact2} and the facts that
\begin{align}
&\int_0^1\rmd r\,r\text{J}_0(qr)=\frac{\text{J}_1(q)}{q}\nonumber \\
\text{and}\quad &\int_1^\infty\rmd r\,r\text{J}_0(kr)\text{J}_0(qr)=\frac{\delta(k-q)}{k}-\frac{1}{k^2-q^2}\left[k\text{J}_0(q)\text{J}_1(k)-q\text{J}_0(k)\text{J}_1(q)\right],
\end{align}
Eq.~\eqref{BCComb1} reduces to the following integral equation for $A(k)$ (interchanging $k\leftrightarrow q$):
\begin{align}
&A(k)\left(k-\frac{\text{Bo }\Omega^2}{\text{Bo}+k^2}\right)=\mathrm{i}\Omega Zb(k)-\Omega^2\int_0^{\infty}\mathrm{d}q\,F(q,k)A(q),\nonumber \\
&\text{where}\quad b(k)=\frac{\text{J}_1(k)}{k}+\frac{1}{\text{Bo}+k^2}\left(\beta\text{J}_0(k)-k\text{J}_1(k)\right),\nonumber \\
& F(q,k)=\frac{q\text{ Bo}}{\text{Bo}+q^2}\left\{\frac{q\text{J}_0(k)\text{J}_1(q)-k\text{J}_0(q)\text{J}_1(k)}{q^2-k^2}+\frac{\text{J}_0(q)}{\text{Bo}+k^2}\left(\beta\text{J}_0(k)-k\text{J}_1(k)\right)\right\}\nonumber \\
&\text{and}\quad\beta = \frac{\sqrt{\text{Bo}}\,\text{K}_1\left(\sqrt{\text{Bo}}\right)}{\text{K}_0\left(\sqrt{\text{Bo}}\right)}.\label{IntEq}
\end{align}
Defining
\begin{align}
\tilde{A}(k)=A(k)\left(k-\frac{\text{Bo }\Omega^2}{\text{Bo}+k^2}\right),\label{AtildeDef}
\end{align}
we obtain the following regularized Fredholm integral equation of the second kind:
\begin{align}
&\tilde{A}(k)+\Omega^2\lim_{\epsilon\rightarrow 0^+}\int_0^\infty\mathrm{d} q\,\frac{\tilde{F}(q,k)}{q-q_{\text{r}}+\mathrm{i}\epsilon}\tilde{A}(q)=\mathrm{i}\Omega Zb(k),\nonumber \\
\text{where}\quad &\tilde{F}(q,k)=\frac{F(q,k)(q^2+\text{Bo})}{(q-q_{\text{i}})(q-\overline{q}_{\text{i}})}.\label{IntEqReg}
\end{align}
Substituting Eq.~\eqref{AtildeDef} into Eqs.~\eqref{PhiHankel} and~\eqref{h1Soln}, we obtain expressions for the velocity potential and wave height:
\begin{subequations}
\begin{align}
\Phi(r,z)&=\lim_{\epsilon\rightarrow 0^+}\int_0^{\infty}\rmd k\,k\frac{\tilde{A}(k)}{k-q_{\text{r}}+\rmi\epsilon}\frac{\text{Bo}+k^2}{(k-q_{\text{i}})(k-\overline{q}_{\text{i}})}\text{J}_0(kr)\rme^{kz},\label{PhiSoln} \\
H(r)&= -\mathrm{i}\Omega\text{ Bo}\lim_{\epsilon\rightarrow 0^+}\int_0^{\infty}\mathrm{d} k\,\frac{k}{(k-q_{\text{i}})(k-\overline{q}_{\text{i}})}\frac{\tilde{A}(k)}{k-q_{\text{r}}+\rmi\epsilon}\left[\text{J}_0(kr)-\frac{\text{K}_0\left(r\sqrt{\text{Bo}}\right)}{\text{K}_0\left(\sqrt{\text{Bo}}\right)}\text{J}_0(k)\right]\nonumber \\
&\phantom{=}+Z\frac{\text{K}_0\left(r\sqrt{\text{Bo}}\right)}{\text{K}_0\left(\sqrt{\text{Bo}}\right)}.\label{h1Soln2}
\end{align}\label{PhiAndHSoln}
\end{subequations}
\markblue{In Appendix~\ref{App:RadCond}, we will show that the $+\rmi\epsilon$ terms in Eqs.~\eqref{IntEqReg} and~\eqref{PhiAndHSoln} are such that the radiation condition~\eqref{RadCond} is satisfied. In Appendix~\ref{App:Local}, we will show that the velocity potential and vertical fluid velocity at the interface have the following behavior near the edge of the disc:
\begin{align}
\Phi(r,0)&=\hat{P}(r)-\frac{\rmi\Omega H^{\prime}(1)}{2\pi}(r-1)^2\log|r-1| +O\left((r-1)^3\log|r-1|\right), 
\nonumber \\
\partial_z\Phi(r,0)&=\rmi\Omega Z+\rmi\Omega H^{\prime}(1)(r-1)_++o(r-1)\quad\text{as}\quad r\rightarrow 1,\label{LocalEqn}
\end{align}
where $\hat{P}(r)$ is $C^2$ at $r=1$ and the notation $\left(\cdot\right)_+$ denotes the positive part. That is,
the vertical fluid velocity has a one-sided ``kink" at the edge of the disc, as it transitions from being prescribed by the disc's motion to satisfying the kinematic and dynamic boundary conditions. This is in contrast to the gravity-wave case ($\text{Bo}=\infty$), for which $\partial_z\Phi(r,0)$ has a discontinuity at the edge of the disc (Appendix~\ref{App:LocalComparison}).
}

To compute the disc's complex oscillation amplitude $Z$ in Eq.~\eqref{ZEqnND}, we require expressions for $\int_0^1 \rmd r\,r\Phi(r,0)$ and $H^{\prime}(1)$, which readily follow from Eq.~\eqref{PhiAndHSoln}: 
\begin{subequations}
\begin{align}
&\int_0^1\rmd r\,r\Phi(r,0)=\lim_{\epsilon\rightarrow 0^+}\int_0^{\infty}\rmd k\,\frac{\tilde{A}(k)}{k-q_{\text{r}}+\rmi\epsilon}\frac{\text{Bo}+k^2}{(k-q_{\text{i}})(k-\overline{q}_{\text{i}})}\text{J}_1(k),\label{PhiInt} \\
&H^{\prime}(1)= -\beta Z+\rmi\Omega\lim_{\epsilon\rightarrow 0^+}\int_0^{\infty}\rmd k\,\frac{k\text{ Bo}}{(k-q_{\text{i}}) (k-\overline{q}_{\text{i}})}\frac{\tilde{A}(k)}{k-q_{\text{r}}+\rmi\epsilon}\left(k\text{J}_1(k)-\beta\text{J}_0(k)\right).\label{hp} 
\end{align}\label{hpPhiInt}
\end{subequations}

In \S~\ref{Sec:Physical} we will compare Eq.~\eqref{PhiInt} with its corresponding expression in the gravity-wave regime, $\text{Bo}=\infty$. The gravity-wave case is a special case of the problem solved herein; specifically, from Eq.~\eqref{DispRel} we have $q_{\text{r}}\sim \Omega^2$ and $q_{\text{i}}\sim \rmi\sqrt{\text{Bo}}$ as $\text{Bo}\rightarrow\infty$. We thus obtain the same integral equation~\eqref{IntEqReg}, with the simplified expressions
\begin{align}
b(k)=\frac{\text{J}_1(k)}{k}\quad\text{and}\quad \tilde{F}(q,k)=\frac{q}{q^2-k^2}\left(q\text{J}_0(k)\text{J}_1(q)-k\text{J}_0(q)\text{J}_1(k)\right)\quad\text{for}\quad\text{Bo}=\infty.\label{IntQuantBoInf}
\end{align}
Equation~\eqref{PhiInt} simplifies as
\begin{align}
\int_0^1\rmd r\,r\Phi(r,0)=\lim_{\epsilon\rightarrow 0^+}\int_0^{\infty}\rmd k\,\frac{\tilde{A}(k)}{k-\Omega^2+\rmi\epsilon}\text{J}_1(k)\quad\text{for}\quad \text{Bo}=\infty.\label{PhiIntBoInf}
\end{align}

\subsection{Numerical method}\label{SSec:Numerical}
We now describe a method for numerically solving the integral equation~\eqref{IntEqReg}. To evaluate the $\epsilon$-limit, we \markblue{recall the Plemelj formula for a function $f(q)$:}
\begin{align}
\lim_{\epsilon\rightarrow 0^+}\int_0^{\infty}\rmd q\,\frac{f(q)}{q-q_{\text{r}}+\rmi\epsilon}=\lim_{\epsilon\rightarrow 0^+}\int_{|q-q_{\text{r}}|>\epsilon,q>0}\rmd q\,\frac{f(q)}{q-q_{\text{r}}}-\rmi\pi f(q_{\text{r}}).\label{PVCalc0}
\end{align}
To evaluate the integral on the right-hand side, we fix some $a > q_{\text{r}}$ and obtain
\begin{align}
\lim_{\epsilon\rightarrow 0^+}\int_{|q-q_{\text{r}}|>\epsilon,q>0}\rmd q\,\frac{f(q)}{q-q_{\text{r}}}&=f(q_{\text{r}})\log\left(\frac{a-q_{\text{r}}}{q_{\text{r}}}\right)+\int_0^a\rmd q\,\frac{f(q)-f(q_{\text{r}})}{q-q_{\text{r}}}\nonumber \\ &\phantom{=}+\int_a^{\infty}\rmd q\,\frac{f(q)}{q-q_{\text{r}}}.\label{PVCalc1}
 \end{align}
Note that both integrals on the right-hand side above are well-defined and can readily be evaluated numerically using the trapezoidal rule. We use the same approach for evaluating the integrals in Eq.~\eqref{hpPhiInt}.

To solve Eq.~\eqref{IntEqReg} numerically, we discretize the interval $0\leq k \leq k_{\text{max}}$ using $N$ non-uniformly spaced points $k_i$, $i=1,\dots,N$, with $N_1$ equispaced points on the interval $[0,q_{\text{r}})$, $N_2$ equispaced points on $[q_{\text{r}},a]$ and the remaining equispaced points on $(a,k_{\text{max}}]$. To resolve the integral accurately, we use a higher density of points in the interval $[0,a]$ than in $[a,k_{\text{max}}]$, since $\tilde{A}(k)$ decays as $k\rightarrow\infty$. The integral in~\eqref{IntEqReg}, rewritten using~\eqref{PVCalc0} and~\eqref{PVCalc1}, is discretized using the non-uniform trapezoidal rule, which yields a system of $N$ equations in the $N$ unknowns $\tilde{A}(k_i)$. This linear system is solved directly using the $\tt backslash$ operator in MATLAB. For the parameter values adopted in this paper, we find that $a=10q_{\text{r}}$, $k_{\text{max}}=200q_{\text{r}}$, $N_1=200$, $N_2=1800$ and $N=5000$ provides sufficient resolution. \markblue{In Appendix~\ref{App:Convergence}, we provide evidence that the numerical algorithm converges, by demonstrating that the approximate error of the numerical solution decreases as the number of points $N$ is increased progressively.}

\subsection{\markblue{Simplification for low frequency}, $\Omega\ll 1$}\label{SSec:LowFreq}

For relatively low forcing frequencies, a more accurate solution of Eq.~\eqref{IntEqReg} can be obtained numerically by using the Liouville--Neumann series: 
\markblue{
\begin{align}
\tilde{A}(k)&=\tilde{A}_0(k)+\tilde{A}_1(k)+\tilde{A}_2(k)+\dots.
\end{align}
}
The leading term is simply 
\begin{align}
\tilde{A}_0(k)=\mathrm{i}\Omega Zb(k),\label{A0Soln}
\end{align}
and the remaining terms can be obtained iteratively using the formula
\begin{align}
\tilde{A}_n(k)=-\markblue{\Omega^2}\lim_{\epsilon\rightarrow 0^+}\int_0^{\infty}\mathrm{d}q\,\frac{\tilde{F}(q,k)}{q-q_{\text{r}}+\mathrm{i}\epsilon}\tilde{A}_{n-1}(q),\quad n\geq 1.\label{LNIter}
\end{align}
The series converges 
\markblue{provided that the spectral radius} of the integral operator in Eq.~\eqref{LNIter} 
is less than \markblue{1}. 
\markblue{This method} can be more accurate than the method presented in \S~\ref{SSec:Numerical} because \markblue{we can use a larger number of points $k=k_i$}: the dense matrix corresponding to the discretization of the integral operator in Eq.~\eqref{IntEqReg} does not need to be constructed, so we are not constrained by the computational memory required to store it. 
In practice we typically use $k_{\text{max}}=2000q_{\text{r}}$ and $N=50000$ points, keeping the values of $a$, $N_1$ and $N_2$ the same as those reported in \S~\ref{SSec:Numerical}.

\section{Experimental setup}\label{Sec:Exp}

\begin{figure}
     \centering
     \includegraphics[width = \linewidth]{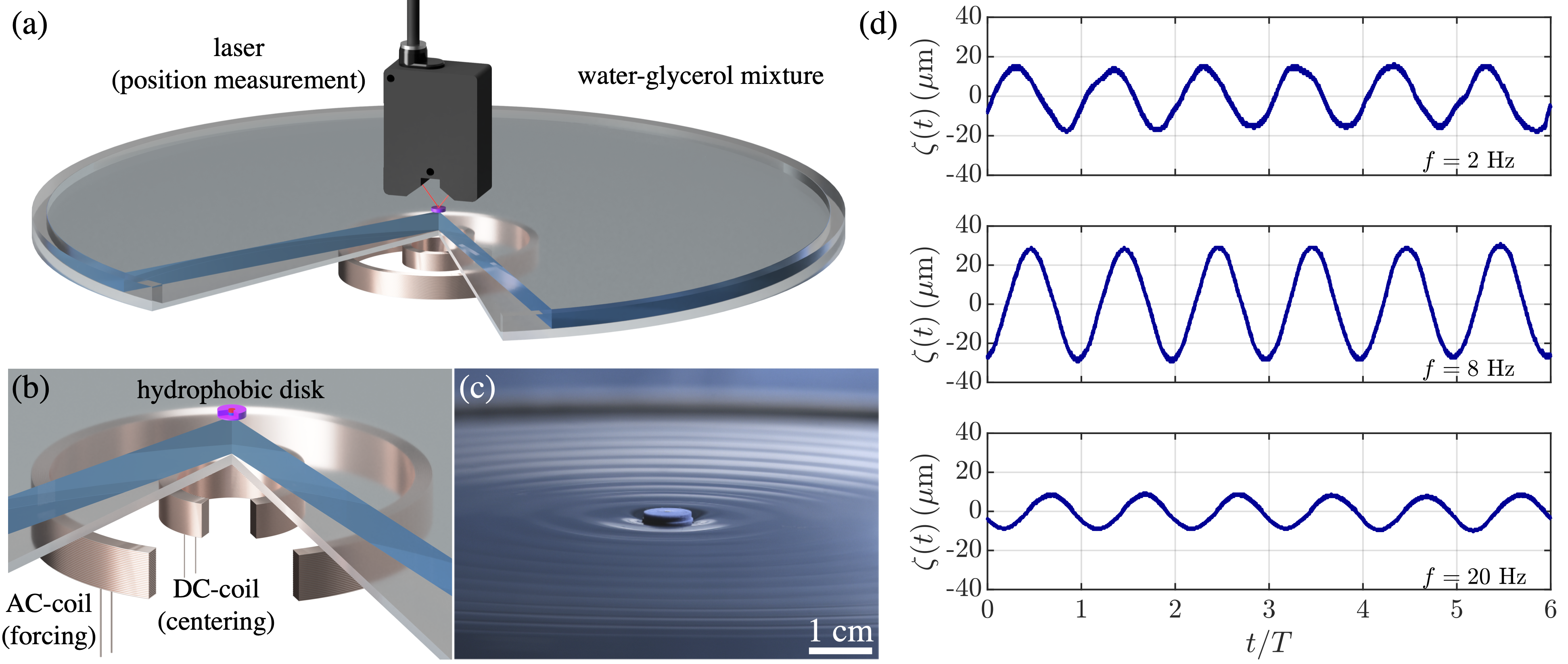}
     \caption{(\textit{a}) A schematic of the experimental setup. The floating disc is oscillated vertically, with its position tracked by a laser position sensor mounted directly above. (b) A zoomed-in view on the floating disc driven to oscillate via an AC coil and aligned with the center of the bath via a DC coil. \textit(c) Experimental image of a disc of radius $R=0.4$ cm driven at $30$ Hz. \markblue{The 1 cm scale was measured at the position of the disc's center}. (\textit{d}) Time traces of the vertical position of a disc of radius $R=0.4$ cm and mass $m=0.081$ g for 6 periods of oscillation. The frequencies shown are 2, 8 and 20 Hz, demonstrating both the non-monotonic amplitude response and increasing phase lag. The water-glycerol mixture has density $\rho = 1.167$ g/cm$^3$, surface tension $\sigma = 66.5$ g/s$^2$, and kinematic viscosity $\nu = 13$ cSt.}
     \label{fig:ExpSetup}
 \end{figure}

The experimental setup is depicted in figure \ref{fig:ExpSetup}(a-c). A centimetric cylindrical disc of radius $R$ and mass $m$ is placed atop a water-glycerol bath of density $\rho$, surface tension $\sigma$, and kinematic viscosity $\nu$. \markblue{The disc is carefully placed at the interface, ensuring that the contact line remains pinned along its bottom perimeter to prevent side-wall wetting, which would induce a visible tilt to the disc in equilibrium.} The discs are made from a commercially available silicone rubber compound (OOMOO 30), which is naturally hydrophobic. The two-part OOMOO mixture is mixed in equal parts (by volume) and then poured into 3D-printed molds fabricated on a Formlabs Form 3+ resin printer. The cast discs are left to sit overnight and then manually removed from the 3D-printed molds. Each disc is designed with a small cylindrical recession on its upper face, wherein a vertically polarized neodymium magnet (SuperMagnetMan D0110-10) is installed, with permanent magnetic dipole moment of $\mathcal{M} = 9$ A$\cdot$cm$^2$. Given the resolution of the 3D-printed molds and variability in casting between individual discs, we estimate the uncertainty on the mass to be $\pm10$\% of the reported values, and the uncertainty on the radius of the object as $\pm0.025$ mm. \markblue{The current casting technique for creating discs determined the smallest disc radius we could test, $R = 0.25$ cm $(\text{Bo}=1.1$). For smaller discs, we found that imperfections in the 3D-printed molds and trapped air bubbles during silicone casting led to noticeable surface defects and mass imbalances.}

The density of each water-glycerol mixture is measured daily with a density meter (Anton Paar DMA 35A). The corresponding surface tension  and kinematic viscosity are estimated from tabulated values \citep{glycerine1963physical, cheng2008formula}. The contact line remains pinned to the bottom circular perimeter of the disc throughout the experiments. The radius of the bath ($R_b=21.5$ cm) is large enough to limit the possible influence of wave reflections off the boundary of the container. The disc is subjected to a sinusoidal vertical forcing of frequency $f = \omega/(2\pi)$, ranging from $0.5-40$ Hz, of the form $F(t) = F_0\cos(\omega t)$.  The details of the forcing are discussed later in this section.

The disc oscillates at the driving frequency and generates outwardly propagating gravity-capillary waves. 
The depth of the bath is fixed at $H_b=1.1$ cm for all experiments. We find that $\tanh(kH_b)\approx 1$ for $f\gtrsim 5$ Hz, where the wavenumber $k$ is determined by the dispersion relation~\eqref{DispRel}, which ensures that the experiments typically operate in the deep-water limit. To minimize ambient contamination of the fluid, the bath is covered with a clear acrylic lid. 
The bath is rinsed with deionized water and dried with Kimtech wipes between each set of experimental trials. The water-glycerol mixture is changed at least every two hours. The values of the experimental parameters and their symbols are presented in Table \ref{tab:kd}.

\begin{table}
  \begin{center}
\def~{\hphantom{0}}
  \begin{tabular}{lccc}
      Parameter  & Symbol   &   Definition & Value (cgs) \\[10pt]
       Radius of disc   & $R$ & - & 0.25-0.75 cm\\
       Radius of bath   & $R_b$ & - & 21.5 cm\\
       Depth of bath   & $H_b$ & - & 1.1 cm\\
        Mass of disc   & $m$ & - & 0.046-0.60 g\\
       Density (water-glycerol mixture)   & $\rho$ & - & 1.141-1.204 g cm$^{-3}$\\
       Kinematic viscosity (water-glycerol mixture)   & $\nu$ & - & 6.9-42 cSt\\
       Surface tension  & $\sigma$ & - & 66-68 dynes cm$^{-1}$ \\
     Magnetic dipole moment & $\mathcal{M}$ &  - & 9 A$\cdot$cm$^2$ \\  
       Gravitational acceleration  & $g$  & - & 981 cm s$^{-2}$\\
       Forcing frequency  & $f$ & - &  0.5-40 Hz\\
       Angular frequency & $\omega$ & $2\pi f$ & 3-250 rad s$^{-1}$ \\
       Dimensionless mass & $M$ & $m/(\rho R^3)$ & 1.1-2.5\\
       Bond number & $\text{Bo}$ & $\rho g R^2/\sigma$ & 1.1-9.7 \\
       Dimensionless angular frequency & $\Omega$ & $\omega\sqrt{R/g}$ & 0.1-5.7
  \end{tabular}
  \caption{Relevant experimental parameters, and their ranges, used throughout the study.}
  \label{tab:kd}
  \end{center}
\end{table}

To measure the vertical position $\zeta(t)$ of the disc, we track it with a laser displacement sensor (Keyence LK-G10) mounted directly above via a 3D-printed mount. A semi-circular cutout in the lid with a diameter of 10 cm allows for unobstructed optical access to the disc. 
A red laser indicator shines on the top surface of the disc to aid with alignment.  The sensor outputs a voltage measurement between $-10$ and 10 V that is proportional to the relative position of the object with sensitivity $0.1$ V/$\mu$m. 

The oscillatory forcing of the disc is provided by a time-dependent magnetic field associated with an alternating-current (AC) coil of mean diameter $10.5$ cm with 275 turns, placed 3.0 cm beneath the fluid surface relative to the center of the coil. The AC coil is driven by an amplifier (Modal Shop 2100E21-400) with frequency and amplitude set by a programmable waveform generator (Siglent SDG1025).  The forcing on the disc arises from the time-dependent vertical magnetic field gradient at the position of the disc: ${F}(t) = F_0\cos(\omega t) = \mathcal{M}B_z\cos(\omega t)$ where $B_z$ is the magnitude of the vertical gradient of the vertical component of the magnetic field.  The strength of the magnetic field gradient is directly proportional to the driving current, and thus the applied force $F(t)$ is also directly proportional to the current. By measuring the magnetic field directly using a linear stage and a magnetometer (Memsic Inc. MMC5983MA) for typical experimental parameters (in the absence of the disc), we estimate a typical vertical forcing 
on the order of one dyne. The AC coil is wound with 26 gauge (0.4 mm) enameled copper wire on a custom 3D-printed spool.

In addition, a smaller concentric direct-current (DC) coil of mean diameter $3.25$ cm with 50 turns is used to keep the disc laterally aligned at the center of the bath, and is located 2.5 cm beneath the fluid surface. The DC coil is independently driven by a constant current supply (TekPower TP3005T) held fixed at $40$ mA.  The DC coil adds an additional constant downward vertical force on the disc. From a similar magnetic field measurement, the magnitude of this force is also estimated to be 
on the order of one dyne, which is two orders of magnitude smaller than the disc's weight, $mg \sim$ 100 dynes. As such, while the DC centering coil force sets a new vertical equilibrium position for the disc, it does so by a negligible amount compared to the disc's own weight, and thus is ignored henceforth. The DC coil is wound with 32 gauge (0.20 mm) enameled copper wire on a custom 3D-printed spool.

To acquire data over a range of frequencies in an automated manner, we interact with an oscilloscope (Siglent SDS 1204X-E) and the waveform generator via a custom MATLAB script. Using the oscilloscope, we record the disc's position (via the voltage output of the sensor) and the voltage drop across the AC coil. The script loops through a vector of driving frequencies up to $40$ Hz, with the frequency order randomized for each trial to avoid any potential hysteresis or other biases. For each frequency, the disc displacement and coil voltage are recorded as a function of time for at least 14 periods of oscillation. We fit the disc's vertical displacement to a sinusoid of form $|\tilde{Z}|\cos(\omega t-\psi_{\text{dc}})$ to obtain the amplitude $|\tilde{Z}|$ and the phase lag between the disc and the coil voltage ($\psi_{\text{dc}}$). Figure \ref{fig:ExpSetup}(d) shows typical time traces for the vertical position $\zeta(t)$ relative to equilibrium for frequencies of 2, 8 and 20 Hz. The disc's vertical position is measured in $\mu$m with typical amplitudes on the order of 10 $\mu \text{m}$. As the frequency is changed, so too do the disc's response amplitude and phase relative to that of the driving. 

The amplifier driving the AC coil has a built-in high-pass filter that influences the magnitude of the voltage supplied to the coil, and in turn the coil current and disc driving force.  In order to account for this effect, we conduct an independent sweep over the range of frequencies tested in experiment ($0.5 \leq \omega/(2\pi) \leq 40$ Hz) with a fixed amplitude input provided to the amplifier.  The $z$- component of the magnetic field is simultaneously measured at the bath's surface via the magnetometer.  We fit the measured gain to the standard functional form for a high-pass filter
\begin{equation}
    G(f) = \frac{f/f_c}{\sqrt{1+\left(f/f_c\right)^2}},
\end{equation}
finding a cut-off frequency of $f_c = 1.60 \pm 0.02$ Hz. To remove the influence of the filter on our data, we compute and report a scaled response amplitude $|Z|=|\tilde{Z}|/G(f)$.  This only has a minimal effect for most data, as $G(f)>0.95$ for $f>5$ Hz, the regime in which the majority of our data is collected.

Through this same measurement we are also able to measure and characterize the frequency-dependent phase lag $\psi_{\text{cf}}$ between the coil voltage and the magnetic field (forcing).  We find that there is an approximately linear relationship given by $\psi_{\text{cf}} = \alpha f$, where $\alpha = -0.0159 \pm 0.0001$ rad$\cdot$Hz$^{-1}$.  Ultimately the phase lag $\psi$ between the disc response and the driving force itself is
\begin{equation}
    \psi = \psi_{\text{dc}} + \psi_{\text{cf}}.
    \label{eq: phaselag_exp}
\end{equation}

For the purposes of normalizing our disc response amplitude curves, we perform reference quasi-static measurements acquired at very low frequencies ($\Omega \equiv \omega\sqrt{R/g} \lesssim 0.2$) for each set of parameters.  These are conducted at $f_0=1.0$ Hz for all but the largest disc ($\text{Bo} = 9.7$) where we use $0.75$ Hz.  These measurements also allow us to validate our phase lag calculation, confirming $\psi\approx0$ for these cases.  Following the same fitting protocol, we can compute the reference amplitude for the quasi-static case $(|\tilde{Z}_\text{0}|)$ from the measured position data.  This value is then also scaled by the gain factor to arrive at the reference static displacement $|Z_0|=|\tilde{Z}_\text{0}|/G(f_0)$.

\section{Comparison between theory and experiment}\label{Sec:ExpTheory}

The two quantities of interest that we will use to compare the theory (\S~\ref{Sec:model}) to the experiment (\S~\ref{Sec:Exp}) are the transmissibility and phase lag. We obtain these quantities by computing the ratio
\begin{align}
\frac{Z}{Z_{0}}&=\frac{2\pi \text{ Bo}^{-1/2}\,\text{K}_1\left(\sqrt{\text{Bo}}\right)/\text{K}_0\left(\sqrt{\text{Bo}}\right) + \pi}{-M\Omega^2-2\pi \text{Bo}^{-1}H^{\prime}(1)/Z+\pi+\rmi\Omega\int_{z=0,r<1}(\Phi/Z)\,\mathrm{d}A}, 
\end{align}
where the expression for $Z_{0}$ is obtained by evaluating Eqs.~\eqref{ZEqnND} and~\eqref{h1Soln} for $\Omega=0$. The transmissibility is $|Z/Z_0|$ and the phase lag is $\psi = -\arg(Z/Z_0)$.

\begin{figure}
     \centering
     \includegraphics[width = 1\linewidth]{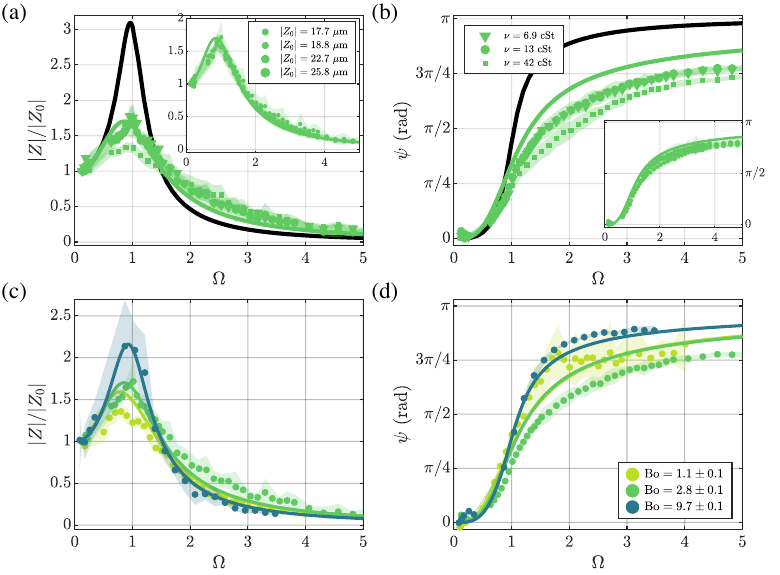}
     \caption{The transmissibility \textit{(a)} and phase lag \textit{(b)} of a vertically driven disc \markblue{(with parameters $\text{Bo}=2.8\pm0.1$ and $M=1.1\pm0.1$)}, with different values of the fluid kinematic viscosity indicted by the different marker shapes. The insets show experiments with a fluid with kinematic viscosity $\nu = 13$ cSt, for different values of the driving amplitude as indicated by circular markers of different sizes. The theoretical predictions for the corresponding dimensionless values $\text{Bo}=2.8$ and $M=1.1$ are shown in the green curves. The black curves indicate the prediction of a theory in which surface tension is neglected ($\text{Bo}= \infty$). The transmissibility \textit{(c)} and phase lag \textit{(d)} of three different discs with the values of $\text{Bo}$ shown, compared to the predictions of the theory (solid curves). \markblue{The dimensionless disc masses are $M =  2.5\pm0.3$, $1.1\pm0.1$ and $1.2\pm0.1$ for the data sets with $\text{Bo} = 1.1\pm0.1$, $2.8\pm0.1$ and $9.7\pm0.1$}, respectively. One standard deviation of the experimental data is shown as the shaded regions. Note that the theory curves in panel (d) largely overlap for $\text{Bo} = 1.1$ and $\text{Bo} = 2.8$. 
     }
     \label{fig:ExpTheory}
\end{figure}

We begin with a disc of radius $R = 0.4$ cm and mass $m = 0.081$ g, which corresponds to $\text{Bo} = 2.8\markblue{\pm0.1}$ and $M = 1.1\markblue{\pm0.1}$, and compare the experimental measurements of the transmissibility and phase lag with our theoretical predictions (figure \ref{fig:ExpTheory}(a,b)). 
The different values of the fluid kinematic viscosity are indicated by the different marker shapes. We also vary the magnitude of the driving force (insets in figure \ref{fig:ExpTheory}(a,b)): since we do not know the precise value of this force, we instead report the value of the disc's static vertical displacement $|Z_0|$. The mean values of the measured quantities are represented by solid markers calculated across six independent trials; the shaded regions denote one standard deviation between these trials. 

Generally, we find good agreement between theory and experiment. The experimental transmissibility is mostly insensitive to the kinematic viscosity of the fluid, except for the most viscous fluid considered ($\nu = 42$ cSt, square markers in figure~\ref{fig:ExpTheory}(a)). The phase lag typically decreases with increasing viscosity (figure~\ref{fig:ExpTheory}(b)). 
While the kinematic viscosity varies by an order of magnitude, the density varies by less than 10\%. It should be noted that while the vertical dynamics of the disc is evidently unaffected by viscosity, the far-field behavior of the resultant wavefield presumably is. 
Prior theoretical work has shown that, within the quasipotential framework, interfacial waves generated by an oscillating point source undergo markedly faster attenuation in the presence of viscosity relative to the inviscid case~\citep{OzaSurfers}. While the far-field expression of the wavefield $H(r)$ in Eq.~\eqref{HAsym2} predicts an algebraic decay of the wavefield, a visual estimate of the wavefield in experiment suggests a faster decay (figure~\ref{fig:ExpSetup}(c)). However, the vertical dynamics of the disc is influenced by the near-field behavior of the wavefield, specifically, the slope of the interface at the contact line and the dynamic pressure underneath the disc, both of which are evidently captured well by our inviscid theory. 

We also note that a theory in which surface tension is neglected (black curves in figure~\ref{fig:ExpTheory}(a,b)) exhibits poor agreement with experiment, in that both the maximum transmissibility and phase lag are overpredicted substantially. {\markblue{Specifically, the gravity-only theory overpredicts the maximum transmissibility by over 10\% for $\text{Bo}\lesssim 150$.}} We thus conclude that accounting for surface tension is necessary to capture the dynamics of discs at the intermediate Bond numbers considered herein. In addition, we find that, for the range of disc amplitudes tested, the transmissibility and phase lag are independent of the driving force amplitude (insets of figure \ref{fig:ExpTheory}(a,b)), indicating that the experiments operate in the linear response regime. Since our linearized theory assumes waves of small amplitude, we expect that substantially larger driving forces on the disc would lead to nonlinear effects that are not captured by our theory.

Finally, in figure \ref{fig:ExpTheory}(c,d) we compare theory and experiment for three different discs, each with distinct values of the Bond number $\text{Bo}$ and dimensionless mass $M$.  For all experiments, the viscosity is fixed at $\nu = 13$ cSt. Generally, we find good agreement between the experimental data and the theoretical predictions for both transmissibility and phase lag. As the Bond number increases, we tend to see an increase in the maximal value of the transmissibility (figure \ref{fig:ExpTheory}(c)) and a corresponding faster increase of the phase lag as the frequency $\Omega$ is increased (figure \ref{fig:ExpTheory}(d)). In analogy with a simple harmonic oscillator, we can see that the inclusion of surface tension appears to add resistance to the driven system, an effect that we quantify in \S~\ref{Sec:Physical}. In Appendix~\ref{App:Extra} (figure~\ref{fig:ExpExtra}), we present additional results with three other discs, displayed alongside the three presented in figure~\ref{fig:ExpTheory}(c,d). We generally find favorable agreement between theory and experiment for all six discs considered.

\section{Physical interpretation of results}\label{Sec:Physical}

\begin{figure}
     \centering
     \includegraphics[width = 1\linewidth]{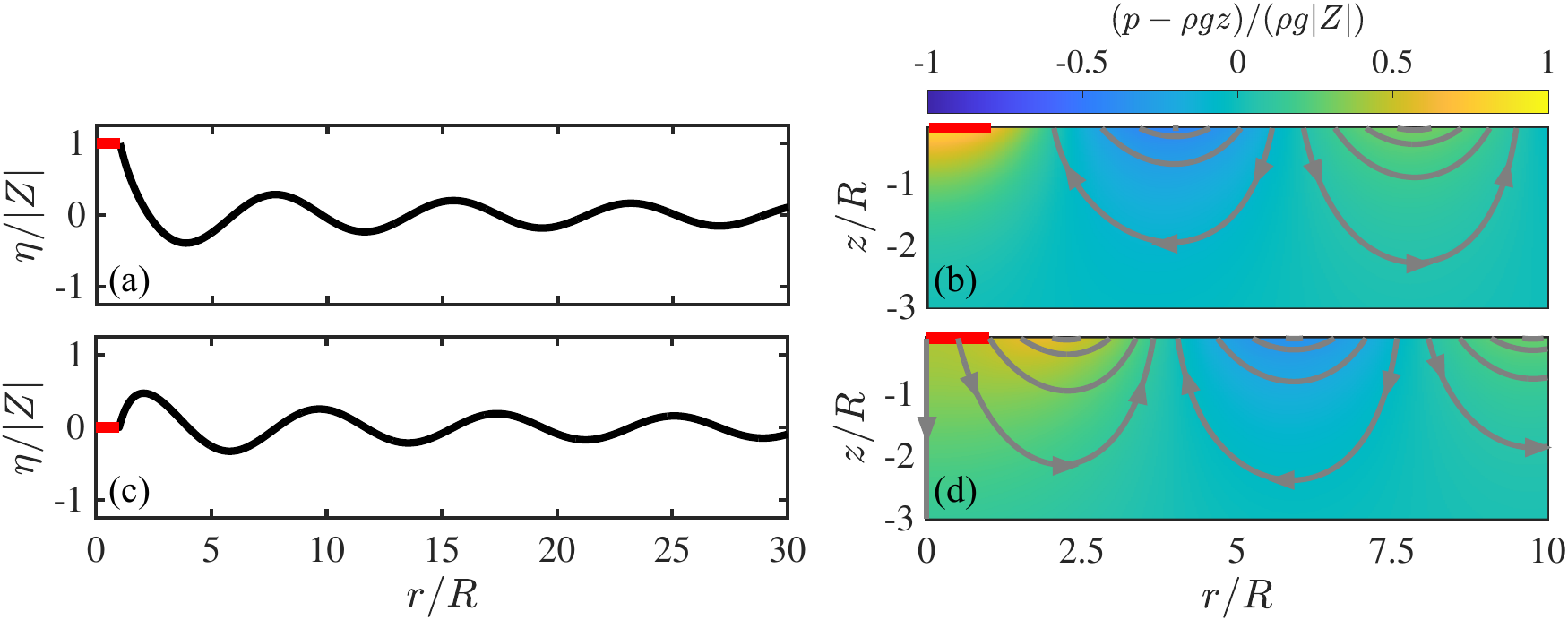}
     \caption{(a) Plot of the wavefield $\text{Re}[H(r)]$ at the times $t=[2n\pi-\arg(Z)]/\omega$, at which the disc (red line) is at its maximum amplitude and has zero velocity, $\zeta=|Z|$ and $\dot{\zeta}=0$. The parameters are $\text{Bo}=3$ and $\Omega=1$. (b) Colormap shows the pressure with the hydrostatic contribution removed, $\text{Im}[\Omega\Phi(r,z)]$, at the same times. Gray lines show streamlines of the flow. Note that the horizontal axis is different to that in panel (a). (c) Plot of the wavefield $-\text{Im}[H(r)]$ at the times $t=[(2n+1/2)\pi-\arg(Z)]/\omega$, at which the disc is moving down with its maximum speed, $\zeta=0$ and $\dot{\zeta}=-\omega |Z|$. (d) Plot of the corresponding pressure field $\text{Re}[\Omega\Phi(r,z)]$.  }
     \label{fig:WavePic}
\end{figure}

In this section, we analyze our theoretical predictions more closely, in order to physically interpret the results presented in figures~\ref{fig:ExpTheory} and~\ref{fig:ExpExtra}. We begin with visualizations of the wavefield, pressure field and flows generated by an oscillating disc with $\text{Bo}=3$ and $\Omega=1$, which are shown in figure~\ref{fig:WavePic} and Supplementary Movie 1. The wavefield $\eta$ is shown in figure~\ref{fig:WavePic}(a,c), and the corresponding pressure field (with the hydrostatic contribution removed) in figure~\ref{fig:WavePic}(b,d). Panels (a,b) show the solution at the instants in the oscillation cycle for which the disc is at its maximum amplitude and has zero velocity, whereas panels (c,d) correspond to when the disc has zero amplitude and is moving downwards with its maximum speed. The oscillatory nature of the wavefield is evident from figure~\ref{fig:WavePic}(a,c), and the wavelength of oscillation is close to the prediction $2\pi/q_{\mathrm{r}}\approx 7.7$ of the gravity-capillary dispersion relation~\eqref{DispRel}. Supplementary Movie 1 shows that the waves radiate outwards from the disc, as required by the radiation condition~\eqref{RadCond}. The oscillatory nature of the pressure field near the surface ($z\lesssim 0$) is evident from figure~\ref{fig:WavePic}(b,d), with relatively low (blue) values of the pressure corresponding to troughs in the wavefield $\eta$ and high (yellow) values to the peaks. The pressure decays with depth $z$, as expected from the form of the solution~\eqref{PhiHankel}. The gray curves indicate streamlines of the flow, which exhibit alternating regions of downwelling and upwelling. 

\begin{figure}
     \centering
     \includegraphics[width = 1\linewidth]{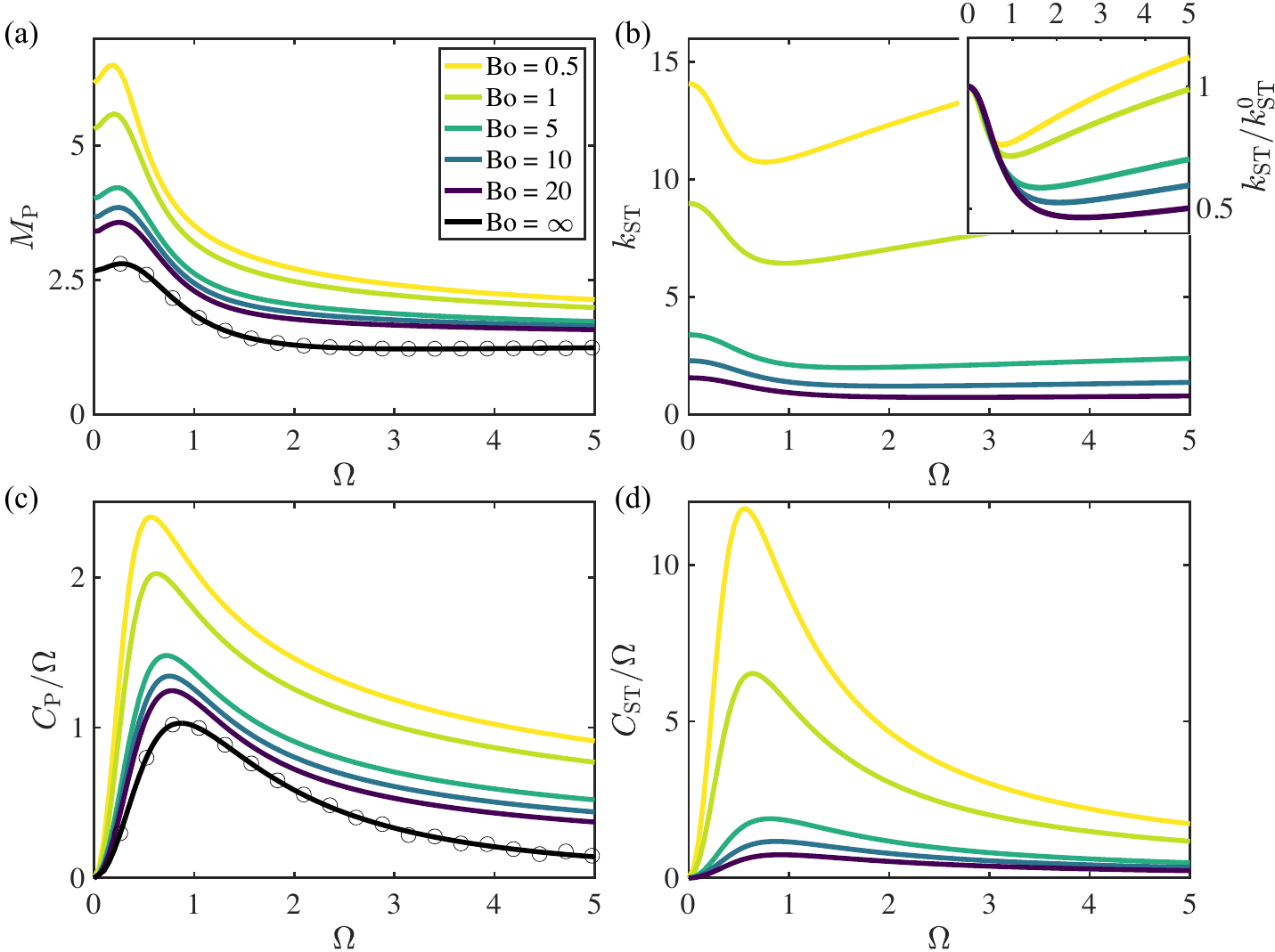}
     \caption{ \textit{(a)} The added mass coefficient $M_{\mathrm{P}}$ of a driven disc as a function of the dimensionless frequency, $\Omega$, for the different Bond numbers indicated in the legend. The black curve corresponds to the gravity-wave regime in which surface tension is neglected, $\mathrm{Bo}=\infty$. \textit{(b)} The dependence of the capillary spring constant $k_{\mathrm{ST}}$ on $\Omega$ for different $\mathrm{Bo}$. The inset shows the ratio of $k_{\mathrm{ST}}$ to the static spring constant $k_{\mathrm{ST}}^0$ defined in Eq.~\eqref{kST0}. \textit{(c)} The resistance (wave-damping) coefficient $C_{\mathrm{P}}/\Omega$ resulting from the force due to dynamic pressure. \textit{(d)} The resistance coefficient $C_{\mathrm{ST}}/\Omega$ resulting from the surface tension force. \markblue{In panels \textit{(a)} and \textit{(c)}, the gravity-wave result from~\citet{Miles1987} is indicated by the circular data points.}}
     \label{fig:QuantPlot}
\end{figure}

We now turn to the added mass $M_{\text{P}}$; damping coefficients $C_{\text{P}}$ and $C_{\text{ST}}$ due to dynamic pressure and surface tension, respectively; and the capillary spring constant $k_{\text{ST}}$, which were defined in Eq.~\eqref{CoeffDef}. Figure~\ref{fig:QuantPlot} shows the dependence of these quantities on the dimensionless forcing frequency $\Omega$ and Bond number $\text{Bo}$. Starting with the added mass $M_{\mathrm{P}}$ in figure~\ref{fig:QuantPlot}(a), we see that, for a given $\text{Bo}$, it increases for small $\Omega$, before reaching a maximum and then decreasing as $\Omega$ is increased further. The added mass evidently decreases with $\mathrm{Bo}$, showing that the inclusion of surface tension results in the disc acquiring a larger effective mass. The added mass in {\markblue{the absence of surface tension ($\text{Bo}=\infty$)}} 
is obtained by evaluating Eq.~\eqref{PhiIntBoInf}, and is shown in the black curve in figure~\ref{fig:QuantPlot}(a). We find that result to be consistent with the numerical results obtained by~\citet[figure 6]{MacCamy1961a} and~\citet{Miles1987}, {\markblue{the latter of which is indicated by the circular data points in figure~\ref{fig:QuantPlot}(a)}}. 

Figure~\ref{fig:QuantPlot}(b) shows the capillary spring constant $k_{\mathrm{ST}}$. Somewhat surprisingly, $k_{\mathrm{ST}}$ decreases at low values of $\Omega$, before reaching a minimum as $\Omega$ is increased progressively and increasing thereafter. The effective spring associated with capillary-wave radiation evidently exhibits different behavior between low and high forcing frequency, softening for the former and stiffening for the latter. As expected, the values of $k_{\mathrm{ST}}$ decrease with $\text{Bo}$, as capillary forces weaken as the Bond number is increased. An alternative comparison between these curves is obtained by plotting the ratio of $k_{\mathrm{ST}}$ to $k_{\text{ST}}^0$, the ``static" capillary spring constant in the absence of oscillatory forcing ($\Omega=0$):
\begin{align}
k_{\text{ST}}^0=\frac{2\pi}{\sqrt{\text{Bo}}}\frac{\text{K}_1\left(\sqrt{\text{Bo}}\right)}{\text{K}_0\left(\sqrt{\text{Bo}}\right)}.
\label{kST0}
\end{align}
The corresponding curves are shown in the inset of figure~\ref{fig:QuantPlot}(b), which appear to nearly overlap for $0\leq \Omega\lesssim 1$, before spreading apart from each other for $\Omega > 1$. The ratio $k_{\mathrm{ST}}/k_{\mathrm{ST}}^0$ evidently decreases with $\text{Bo}$ for moderate to large forcing frequencies ($\Omega > 1.5$). By splitting the capillary spring constant into static and dynamic components, $k_{\mathrm{ST}}=k_{\mathrm{ST}}^0+k_{\mathrm{ST}}^{\text{dyn}}$, we thus conclude that dynamic capillary spring forces associated with $k_{\mathrm{ST}}^{\text{dyn}}$ become increasingly dominant as $\text{Bo}$ is decreased and $\Omega$ is increased.

Figure~\ref{fig:QuantPlot}(c,d) shows the dependence of the damping coefficients on $\Omega$ and $\text{Bo}$. We plot $C_{\mathrm{P}}/\Omega$ and $C_{\mathrm{ST}}/\Omega$, as the former is sometimes referred to as the ``radiation resistance" in the literature~\citep{Miles1987}. It is evident that, for a given Bond number, both quantities vanish in the low-frequency limit and increase with $\Omega$ for relatively small values of $\Omega$, before attaining a maximum and then decreasing to zero as $\Omega$ is increased further. Moreover, both sources of damping decrease with $\text{Bo}$, which explains the observation in both theory and experiment that the disc's oscillation amplitude increases with $\text{Bo}$ (figures~\ref{fig:ExpTheory}(c) and~\ref{fig:ExpExtra}). 

By comparing figure~\ref{fig:QuantPlot}(c) and figure~\ref{fig:QuantPlot}(d), we observe that $C_{\text{ST}}/\Omega$ is typically larger than $C_{\mathrm{P}}/\Omega$ for $\Omega\approx 1$ and $\text{Bo} \leq 5$, indicating that surface tension damping is dominant over pressure damping for the parameter regime in which the experiments typically operate. As the Bond number is increased progressively, capillary forces weaken and pressure forces begin to dominate, with $C_{\mathrm{ST}}/\Omega\rightarrow 0$ as $\text{Bo}\rightarrow\infty$ but $C_{\mathrm{P}}/\Omega$ approaching the gravity-wave result given by the black curve in figure~\ref{fig:QuantPlot}(c). We find that result to be consistent with the calculations of~\citet[figure 7]{MacCamy1961a} and~\citet{Miles1987}, {\markblue{the latter of which is indicated by the circular data points in figure~\ref{fig:QuantPlot}(c)}}.

\begin{figure}
\begin{center}
\includegraphics[width=1\textwidth]{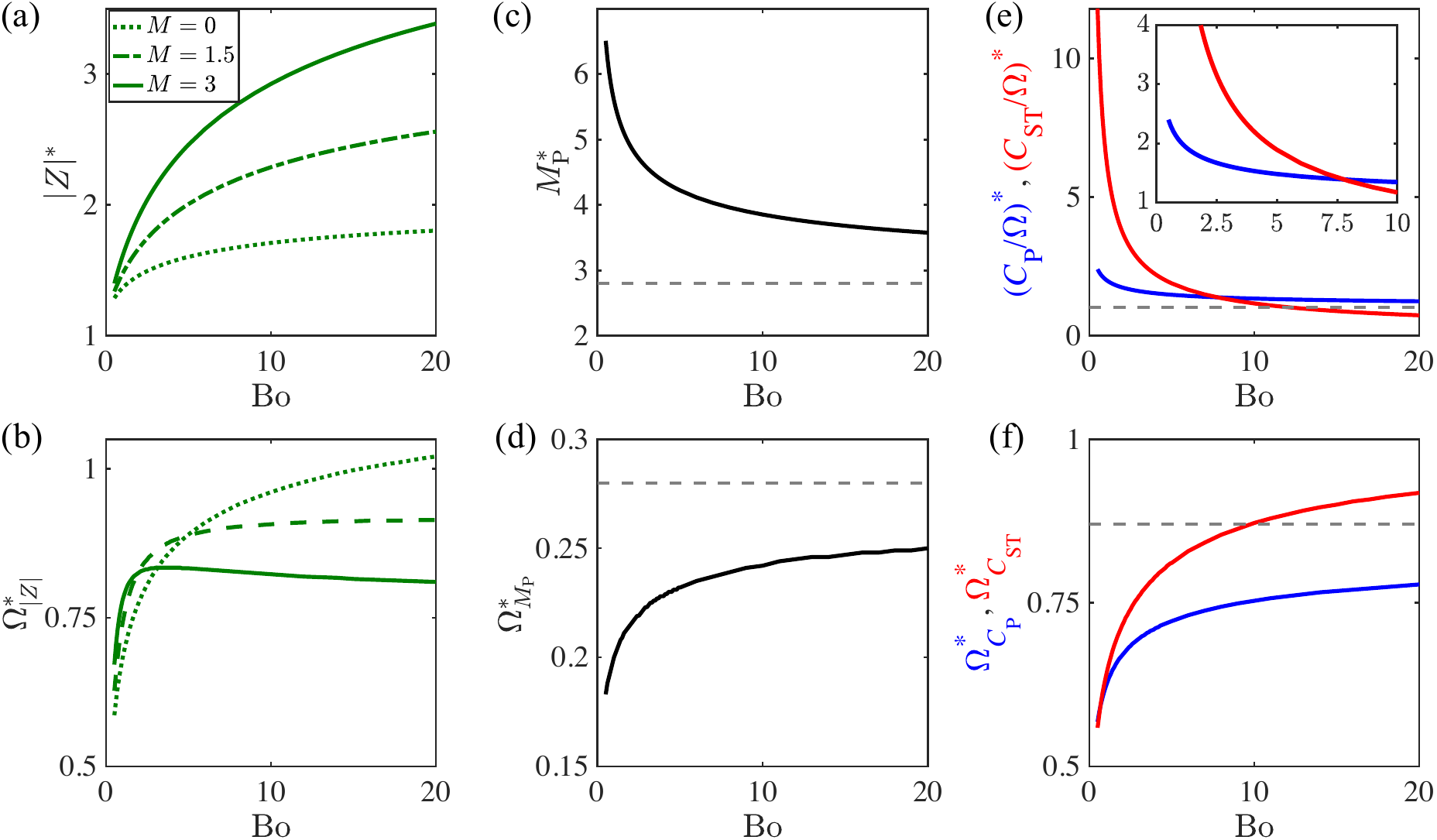}
\caption{(a) Dependence of the maximum oscillation amplitude of the disc $|Z|^*$ on $\text{Bo}$, for three different values of the disc's dimensionless mass $M$. (b) Dependence of the corresponding forcing frequency $\Omega^*_{|Z|}$ at which the disc attains its maximum oscillation amplitude, for the same values of $M$. (c) The maximum added mass $M_{\mathrm{P}}^*$. The dashed line indicates the gravity-wave  result ($\mathrm{Bo}=\infty$). (d) The frequency $\Omega_{M_{\mathrm{P}}}^*$ for which the added mass $M_{\mathrm{P}}$ attains its maximum value. The dashed line corresponds to $\mathrm{Bo}=\infty$. (e) The maximum values of the wave resistances $(C_{\mathrm{P}}/\Omega)^*$ (blue) and $(C_{\mathrm{ST}}/\Omega)^*$ (red). The inset zooms into the region where the two quantities intersect. The dashed line indicates the $\mathrm{Bo}=\infty$ result for $C_{\mathrm{P}}/\Omega$. (f) The frequencies $\Omega_{C_{\mathrm{P}}}^*$ (blue) and $\Omega_{C_{\mathrm{ST}}}^*$ (red) for which the wave resistances attain their maximum values. The dashed line indicates the $\mathrm{Bo}=\infty$ result for $\Omega_{C_{\mathrm{P}}}^*$.} \label{fig:MaxQuants}
\end{center} 
\end{figure}

Each of the discs considered in the experiments (figures~\ref{fig:ExpTheory} and~\ref{fig:ExpExtra}) exhibits a frequency $\Omega_{|Z|}^*$ for which its oscillation amplitude $|Z|$ is largest, $|Z|=|Z|^*$. In figure~\ref{fig:MaxQuants}(a,b) we plot the theoretically predicted values of the maximum amplitude and corresponding frequency as a function of Bond number, for three different values of the dimensionless mass $M$. It is evident that the maximum amplitude increases with Bond number, as observed in the experiments (figures~\ref{fig:ExpTheory} and~\ref{fig:ExpExtra}). For a given Bond number, the maximum amplitude increases with $M$. The dependence of the corresponding frequency on $\text{Bo}$ and $M$ is more complicated; while it increases monotonically with $\text{Bo}$ for a massless disc ($M=0$, dotted curve in figure~\ref{fig:MaxQuants}(b)), the behavior is non-monotonic for a more massive disc ($M=3$, solid curve in figure~\ref{fig:MaxQuants}(b)). 
We observe that, for the range of Bond numbers corresponding to the experiments, $1\leq \text{Bo}\leq 10$, the frequency corresponding to the maximum oscillation amplitude is in the relatively small interval $0.7\leq \Omega_{|Z|}^*\leq 1$ across all three values $M$, indicating that changes to the disc's mass and size have a relatively small effect on $\Omega_{|Z|}^*$.

Similarly, we have observed that the added mass $M_{\mathrm{P}}$ (figure~\ref{fig:QuantPlot}(a)), dynamic pressure resistance $C_{\mathrm{P}}/\Omega$ (figure~\ref{fig:QuantPlot}(c)) and surface tension resistance $C_{\mathrm{ST}}/\Omega$ (figure~\ref{fig:QuantPlot}(d)) exhibit maxima at nonzero values of $\Omega$. Figure~\ref{fig:MaxQuants}(c--f) shows the maximum values of these quantities, and the forcing frequencies for which these maxima are attained, as functions of $\text{Bo}$ for the range $0.5\leq\text{Bo}\leq 20$. We find that the maximum added mass $M_{\mathrm{P}}^*$ decreases monotonically with $\text{Bo}$ (figure~\ref{fig:MaxQuants}(c)), and the corresponding frequency $\Omega_{M_{\mathrm{P}}}^*$ increases weakly with $\text{Bo}$, remaining in the range $0.2\leq \Omega_{M_{\mathrm{P}}}^*\leq 0.25$ (figure~\ref{fig:MaxQuants}(d)). The values of the maximum wave resistance associated with dynamic pressure $(C_\mathrm{P}/\Omega)^*$ and surface tension $(C_{\mathrm{ST}}/\Omega)^*$ also decrease with $\mathrm{Bo}$, as shown in figure~\ref{fig:MaxQuants}(e). Note that the wave resistance associated with surface tension (dynamic pressure) is dominant for small (large) $\text{Bo}$, with a crossover between the two occurring at $\text{Bo}\approx 8$ (figure~\ref{fig:MaxQuants}(e) inset). The frequencies $\Omega_{C_{\mathrm{P}}}^*$ and $\Omega_{C_{\mathrm{ST}}}^*$ corresponding to both maxima evidently increase with $\text{Bo}$, remaining within the interval $0.5\leq \Omega\leq 1$ for the range of $\text{Bo}$ considered herein.


\subsection{Asymptotic expressions for physical quantities as $\Omega\rightarrow 0$}\label{SSec:Asym}

\begin{figure}
\begin{center}
\includegraphics[width=1\textwidth]{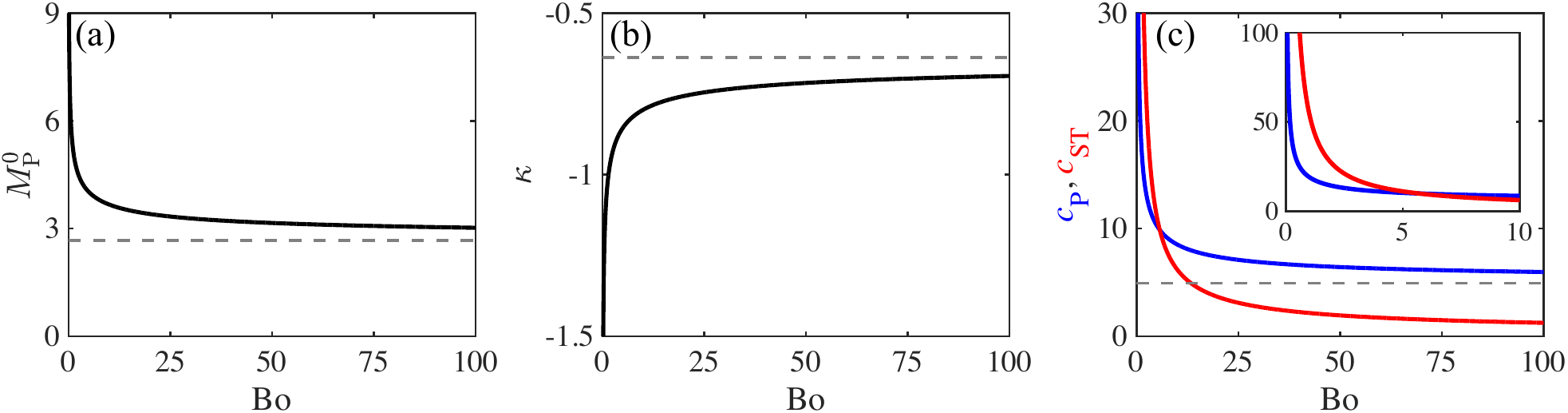}
\caption{Physical quantities in the low-frequency limit $\Omega\rightarrow 0$. (a) Added mass $M_{\text{P}}^0$, as given by Eq.~\eqref{MP0}. The dashed line indicates the value $8/3$, the asymptotic value in the limit $\text{Bo}\rightarrow\infty$ as given by Eq.~\eqref{MP0BoInf}. (b) Capillary spring constant correction $\kappa$, as given by Eq.~\eqref{kappaST}. The dashed line indicates the value $-2/\pi$, the asymptotic value in the limit $\text{Bo}\rightarrow\infty$ (see Eq.~\eqref{kappaBoInf}). (c) Damping coefficients due to dynamic pressure $c_{\text{P}}$ (blue) and surface tension $c_{\text{ST}}$ (red), as given by Eqs.~\eqref{cP} and~\eqref{cST}, respectively. The dashed line corresponds to the value $\pi^2/2$, the asymptotic value of $c_{\text{P}}$ in the limit $\text{Bo}\rightarrow\infty$ (see Eq.~\eqref{cPBoInf}). The inset zooms into the low $\text{Bo}$ region, which shows that the crossover occurs around $\text{Bo}\approx 6$.} \label{fig:Asymptotics}
\end{center} 
\end{figure}

While the values of the added mass, capillary spring constant and damping coefficients shown in figure~\ref{fig:QuantPlot} were obtained by numerically solving the integral equation~\eqref{IntEqReg} for $\tilde{A}(k)$, analytical expressions for these quantities can be obtained in the limit of low forcing frequency ($\Omega\rightarrow 0$) using the leading-order solution $\tilde{A}_0(k)$ in Eq.~\eqref{A0Soln}. Starting with $M_{\text{P}}$, we substitute $\tilde{A}_0(k)$ for $\tilde{A}(k)$ in Eq.~\eqref{PhiInt} and use the facts that $q_{\text{r}}\rightarrow \Omega^2$ and $q_{\text{i}}\rightarrow \rmi\sqrt{\text{Bo}}$ as $\Omega\rightarrow 0$. We thus obtain
\begin{align}
\lim_{\Omega\rightarrow 0}M_{\text{P}}&\equiv M_{\text{P}}^0= 2\pi\int_0^\infty\rmd k\,\frac{b(k)\text{J}_1(k)}{k}.\label{LowOmM}
\end{align}
As detailed in Appendix~\ref{App:Asymptotics}, by substituting the expression for $b(k)$ from Eq.~\eqref{IntEq} into the above integral, we obtain the explicit formula
\begin{align}
M_{\text{P}}^0&=\frac{8}{3}+\beta\left[\frac{\pi^2}{\text{Bo}}\text{I}_0\left(\sqrt{\text{Bo}}\right)\text{I}_1\left(\sqrt{\text{Bo}}\right)-\frac{32}{9}{}_2F_3\left(1,2;\frac{3}{2},\frac{5}{2},\frac{5}{2};\text{Bo}\right)\right]+\frac{\pi^2}{\sqrt{\text{Bo}}}\text{I}_1^2\left(\sqrt{\text{Bo}}\right)\nonumber \\
&\phantom{=}-\frac{8}{3}{}_2F_3\left(1,1;\frac{1}{2},\frac{3}{2},\frac{5}{2};\text{Bo}\right),\label{MP0}
\end{align}
where ${}_pF_q$ is the generalized hypergeometric function, {\markblue{and we recall that $\text{I}_n$ is the modified Bessel function of the first kind of order $n$}}. Figure~\ref{fig:Asymptotics}(a) shows that this quantity is monotonically decreasing in $\text{Bo}$, which is consistent with the behavior observed in figure~\ref{fig:QuantPlot}(a). Equation~\eqref{MP0} simplifies in the limits of small and large $\text{Bo}$:
\begin{subequations}
\begin{alignat}{2}
M_{\text{P}}^0&\sim -\frac{1}{2\gamma+\log(\text{Bo}/4)}\left(\frac{\pi^2}{\sqrt{\text{Bo}}}-\frac{64}{9}\right) \quad&&\text{as}\quad\text{Bo}\rightarrow 0,\label{MP0Bo0} \\
M_{\text{P}}^0&\sim\frac{8}{3}+\frac{4}{\sqrt{\text{Bo}}}
\quad&&\text{as}\quad\text{Bo}\rightarrow\infty,\label{MP0BoInf}
\end{alignat}\label{MP0Asym}
\end{subequations}
where $\gamma= 0.577\ldots$ is the Euler-Mascheroni constant. 
The constant term $8/3$ in Eq.~\eqref{MP0BoInf} was given by~\citet[\S102, 4$^{\circ}$]{Lamb1932}, and it represents the added mass of a disc radiating gravity waves in the low-frequency limit. The added mass in the capillary-wave regime, as given by Eq.~\eqref{MP0Bo0}, evidently diverges as $\text{Bo}\rightarrow 0$. 

Similarly, from~\eqref{hp} we have
\begin{align}
k_{\text{ST}}&\sim k_{\text{ST}}^0\left(1+\kappa\Omega^2\right)\quad\text{as}\quad\Omega\rightarrow 0,\quad
\text{where}\quad\kappa\equiv \frac{\text{Bo}}{\beta}\int_0^{\infty}\rmd k\,\frac{b(k)}{\text{Bo}+k^2}\left[k\text{J}_1(k)-\beta\text{J}_0(k)\right].\label{LowOmk}
\end{align}
The expression for the low-frequency correction $\kappa$ is quite unwieldy and is thus deferred to Eq.~\eqref{kappaST}, but is plotted in figure~\ref{fig:Asymptotics}(b). It is evident that $\kappa$ is always negative, so the effective spring appears to soften at low frequency, a finding that is consistent with the behavior observed in figure~\ref{fig:QuantPlot}(b). As shown in Appendix~\ref{App:Asymptotics}, the expression for $\kappa$ simplifies in the limits of small and large $\text{Bo}$:
\begin{subequations}
\begin{alignat}{2}
\kappa&\sim \frac{\pi}{2(2 \gamma + \log(\mathrm{Bo}/4)) \sqrt{\mathrm{Bo}}}\quad &&\text{as}\quad \text{Bo}\rightarrow 0,\label{kappaBo0} \\
\kappa&\sim -\frac{2}{\pi}  \quad &&\text{as}\quad \text{Bo}\rightarrow\infty.\label{kappaBoInf}
\end{alignat}\label{kappaAsym}
\end{subequations}
That is, the correction $\kappa$ diverges in the low-$\mathrm{Bo}$ limit~\eqref{kappaBo0} but approaches a constant in the high-$\mathrm{Bo}$ limit~\eqref{kappaBoInf}. 

We proceed by assessing the behavior of the damping coefficients $C_{\text{P}}$ and $C_{\text{ST}}$ in the limit $\Omega\rightarrow 0$. Starting with the pressure damping coefficient $C_{\text{P}}$, we use Eq.~\eqref{PhiInt} to obtain
\begin{align}
C_{\text{P}}\sim c_{\text{P}}\Omega^3\quad\text{as}\quad \Omega\rightarrow 0,\quad
\text{where}\quad c_{\text{P}}
=-\lim_{\Omega\rightarrow 0}\frac{2\pi}{\Omega^2}\text{ Im }\left[\lim_{\epsilon\rightarrow 0^+}\int_0^{\infty}\rmd k\,\frac{b(k)}{k-\Omega^2+\rmi\epsilon}\text{J}_1(k)\right].
\end{align}
Using Eq.~\eqref{PVCalc0} and the fact that $\text{J}_1(x)\sim x/2$ as $x\rightarrow 0$, we obtain
\begin{align}
c_{\text{P}}=\lim_{\Omega\rightarrow 0}\frac{2\pi^2}{\Omega^2}b(\Omega^2)\text{J}_1(\Omega^2)=\pi^2\left(\frac{1}{2}+\frac{1}{\sqrt{\text{Bo}}}\frac{\text{K}_1\left(\sqrt{\text{Bo}}\right)}{\text{K}_0\left(\sqrt{\text{Bo}}\right)}\right),\label{cP}
\end{align}
which simplifies in the limits of small and large $\text{Bo}$:
\begin{subequations}
\begin{alignat}{2}
c_{\text{P}} &\sim -\frac{2\pi^2}{\left[2\gamma+\log\left(\text{Bo}/4\right)\right]\text{Bo}}\quad &&\text{as}\quad \text{Bo}\rightarrow 0,\label{cPBo0} \\
c_{\text{P}} &\sim \frac{\pi^2}{2} \quad &&\text{as}\quad \text{Bo}\rightarrow\infty.\label{cPBoInf}
\end{alignat}\label{cPAsym}
\end{subequations}
Note that $c_{\text{P}}$ diverges in the capillary-wave limit $\text{Bo}\rightarrow 0$, but approaches a constant in the gravity-wave limit $\text{Bo}\rightarrow\infty$. This constant was previously calculated by~\citet{Miles1987}. 

The calculation for the surface tension damping coefficient $C_{\text{ST}}$ proceeds in a similar fashion. Using~\eqref{hp} we obtain
\begin{align}
C_{\text{ST}}&\sim c_{\text{ST}}\Omega^3\quad\text{as}\quad \Omega\rightarrow 0,\nonumber \\
\text{where}\quad c_{\text{ST}}&=\lim_{\Omega\rightarrow 0}\frac{2\pi}{\Omega^2}\text{Im}\left[\lim_{\epsilon\rightarrow 0^+}\int_0^{\infty}\rmd k\,\frac{k}{k^2+\text{Bo}}\frac{b(k)}{k-\Omega^2+\rmi\epsilon}\left(k\text{J}_1(k)-\beta\text{J}_0(k)\right)\right]\nonumber \\
&=2\pi^2\left(\frac{1}{2}+\frac{1}{\sqrt{\text{Bo}}}\frac{\text{K}_1\left(\sqrt{\text{Bo}}\right)}{\text{K}_0\left(\sqrt{\text{Bo}}\right)}\right)\frac{1}{\sqrt{\text{Bo}}}\frac{\text{K}_1\left(\sqrt{\text{Bo}}\right)}{\text{K}_0\left(\sqrt{\text{Bo}}\right)},\label{cST}
\end{align}
and in the last line we again use~\eqref{PVCalc0} and the fact that $\text{J}_0(x)\sim 1$ as $x\rightarrow 0$. We thus obtain
\begin{subequations}
\begin{alignat}{2}
c_{\text{ST}} &\sim \frac{8\pi^2}{\left[2\gamma+\log\left(\text{Bo}/4\right)\right]^2\text{Bo}^2}\quad &&\text{as}\quad \text{Bo}\rightarrow 0,\label{cSTBo0} \\
c_{\text{ST}} &\sim \frac{\pi^2}{\sqrt{\text{Bo}}} \quad &&\text{as}\quad \text{Bo}\rightarrow\infty.\label{cSTBoInf}
\end{alignat}\label{cSTAsym}
\end{subequations}
By comparing Eqs.~\eqref{cPBo0} and~\eqref{cSTBo0}, it is evident that $c_{\text{ST}}$ diverges in the capillary-wave limit $\text{Bo}\rightarrow 0$ faster than $c_{\text{P}}$. We observe from Eq.~\eqref{cSTBoInf} that $c_{\text{ST}}$ decays to zero in the gravity-wave limit $\text{Bo}\rightarrow\infty$, as expected. Taken together, the asymptotic behaviors of both $c_{\text{P}}$ and $c_{\text{ST}}$ are evident in figure~\ref{fig:Asymptotics}(c): $c_{\text{ST}}$ is clearly dominant at low $\text{Bo}$, but there is a crossover (highlighted in the inset) at $\text{Bo}\approx 6$, and $c_{\text{P}}$ dominates at large $\text{Bo}$.

\section{Conclusion}\label{sec:conc}

We have presented herein a combined theoretical and experimental study of the vertical dynamics of a floating disc that is driven periodically in the vertical direction. The disc generates a gravity-capillary wavefield, which we assume to be linear and inviscid in our theoretical model for the sake of simplicity. The solution is determined by three dimensionless parameters: the dimensionless forcing frequency $\Omega$, Bond number $\text{Bo}$ and disc mass $M$, as defined in Eq.~\eqref{NDimParam}. Our work generalizes prior work by~\citet{MacCamy1961a} and~\citet{Miles1987} who neglected surface tension and computed the added mass and damping coefficients of a heaving disc that generates pure gravity waves. 

The mathematical problem that we solve is the linear elliptic boundary value problem in Eq.~\eqref{BathEqnND} for the velocity potential $\Phi(r,z)$, which satisfies the no-penetration boundary condition on the disc and the kinematic and dynamic boundary conditions on the free surface (figure~\ref{fig:schematic}). The velocity potential is coupled to the unknown wave height $H(r)$ through the free surface conditions.  We recast the problem using the Hankel transform and obtain a 1D integral equation~\eqref{IntEqReg} for the unknown amplitude $\tilde{A}(k)$, which we solve numerically. Given $\tilde{A}(k)$, we may recover the wavefield and the pressure, examples of which are shown for one set of parameters in figure~\ref{fig:WavePic} and Supplementary Movie 1. We also obtain the forces on the disc due to dynamic pressure and surface tension using Eq.~\eqref{hpPhiInt}, and thus the (complex) vertical response of the disc using Eq.~\eqref{ZEqnND}. As we assume the flow to be inviscid, the disc's motion is damped only by its radiating waves. The vertical response of the disc is compared to that measured in laboratory experiments, in which a floating disc is driven to oscillate by an applied magnetic field (figure~\ref{fig:ExpSetup}). 

The agreement between theory and experiment is {\markblue{good}}, 
as shown in figure~\ref{fig:ExpTheory} and figure~\ref{fig:ExpExtra} for six different discs with Bond numbers between 1 and 10. The complex amplitude of the disc increases with $\Omega$ before reaching a maximum at a critical value $\Omega_{|Z|}^*$ and then decreasing to zero as $\Omega\rightarrow\infty$ (figure~\ref{fig:ExpTheory}(a,c)). The disc oscillates in phase with the applied forcing for low $\Omega$, and tends towards out-of-phase oscillations in the high-frequency limit $\Omega\rightarrow\infty$ (figure~\ref{fig:ExpTheory}(b,d)). Capillary forces play an essential role, as evidenced by the inability of a theory that neglects surface tension to match the experimental data (black curves in figure~\ref{fig:ExpTheory}(a,b)).  The agreement between theory and experiment is poorest for the most viscous fluid (squares in figure~\ref{fig:ExpTheory}(a,b)) but is still adequate, indicating that viscous effects eventually play a measurable role in determining the dynamic response of the disc. \markblue{Specifically, for $\Omega=1$ and for the least viscous fluid tested ($\nu = 6.9$ cSt), the viscous boundary layer is an order of magnitude smaller than the disc radius, $\sqrt{\nu/\omega}/R\approx 0.1$. While the boundary layer is roughly four times smaller than the disc radius for the most viscous fluid tested ($\nu=42$ cSt), it becomes comparable to the disc radius if the viscosity is increased to $\approx 800$ cSt, a value beyond which we expect our inviscid theory to be fully inadequate.}

To interpret our results physically, we use the theory to compute the disc’s added mass $M_{\text{P}}$, capillary spring constant $k_{\text{ST}}$, and wave damping coefficients $C_{\text{P}}/\Omega$ and $C_{\text{ST}}/\Omega$, which are shown in figure~\ref{fig:QuantPlot}.
We find that both the added mass $M_{\text{P}}$ and wave damping due to pressure $C_{\text{P}}/\Omega$ increase as surface tension becomes more dominant (i.e. as $\text{Bo}$ decreases), as is clear by comparing the colored curves in figure~\ref{fig:QuantPlot}(a,c) with the gravity-wave result ($\text{Bo}=\infty$, black curve). The gravity-wave result computed using our method agrees with prior computations in the literature~\citep{MacCamy1961a,Miles1987}. Moreover, the contact line on the disc induces a wave damping due to surface tension, which is dominant at low $\text{Bo}$ but weakens as the Bond number is increased progressively (figure~\ref{fig:QuantPlot}(d)). The contact line also induces a spring-like vertical force on the disc with coefficient $k_{\text{ST}}$; as shown in figure~\ref{fig:QuantPlot}(b), the spring effectively softens at low $\Omega$, before stiffening as $\Omega$ is increased progressively. The maximum oscillation amplitude of the disc $|Z|^*$ is found to increase with both the disc’s mass $M$ and Bond number $\text{Bo}$ (figure~\ref{fig:MaxQuants}(a)). On the other hand, the maximum added mass $M_{\text{P}}^*$ and wave damping coefficients $(C_{\text{P}}/\Omega)^*$ and $(C_{\text{ST}}/\Omega)^*$ all decrease with $\text{Bo}$ (figure~\ref{fig:MaxQuants}(c,e)).

The low-frequency ($\Omega\rightarrow 0$) behavior of the added mass, spring constant and wave damping coefficients is given in \S~\ref{SSec:Asym}. Of particular note are the expressions for the added mass $M_{\text{P}}^0$ and $c_{\text{P}}$ in Eqs.~\eqref{MP0} and~\eqref{cP}, respectively, which generalize and extend the corresponding results of~\citet[\S102, 4$^{\circ}$]{Lamb1932} and~\citet{Miles1987}, respectively, by accounting for surface tension. We find that the added mass $M_{\text{P}}^0$ and low-frequency correction to the spring constant $\kappa$ both diverge as $\text{Bo}\rightarrow 0$, but approach constant values in the gravity-wave regime $\text{Bo}\rightarrow\infty$, behaviors that are evident from figure~\ref{fig:Asymptotics}(a,b) and that we quantify in Eqs.~\eqref{MP0Asym} and~\eqref{kappaAsym}, respectively. Similarly, the leading-order coefficients of the wave damping due to pressure ($c_{\text{P}}$) and surface tension ($c_{\text{ST}}$) diverge as $\text{Bo}\rightarrow 0$ but the latter diverges faster, as is evident by comparing Eqs.~\eqref{cPBo0} and~\eqref{cSTBo0}. However, there is a crossover between the two quantities at $\text{Bo}\approx 6$ (figure~\ref{fig:Asymptotics}(c), inset), indicating that wave damping due to surface tension (dynamic pressure) is dominant at relatively low (high) $\text{Bo}$.

One limitation of the theory presented herein is that viscosity is neglected. The theory does an adequate job of capturing the vertical response of a disc even in the most viscous fluid considered (squares in figure~\ref{fig:ExpTheory}(a,b)), because the disc’s response is driven by local pressure and surface tension forces which are evidently only weakly affected by viscosity. However, the theory fails to capture the far-field viscous decay of the wavefield and flows. \markblue{In a future study, one could solve the quasipotential model for surface waves~\citep{Lamb1932,Dias2008}, valid in the regime of weak viscosity, and compare the predictions of that theory to the waves and flows obtained in experiment.} 
The second limitation of the theory is that it is linearized and thus only valid if the amplitude of the disc’s oscillation is small compared to the wavelength. A large applied force $F_0$ would lead to large oscillation amplitudes, a regime in which nonlinear effects would become relevant. 

In future work, one could extend the formalism presented herein to model pitching motion of a disc, or a combined heaving and pitching motion. In order to understand and model the self-propulsion or spinning of floating objects like the capillary surfers and spinners described in \S~\ref{Sec:Intro}, one would have to extend the formalism presented herein to account for non-axisymmetric waves and flows~\citep{ho2023capillary,Barotta2023,Harris2025Review}. While the results presented herein elucidate the dependence of the added mass, capillary spring constant and wave damping coefficients on the size of the disc, as quantified by the Bond number, all of those physical quantities presumably depend on the shape of the object too. Elucidating the influence of geometry on the associated waves and flows is a potential future avenue of research that could inform the design of new microrobots that self-propel while oscillating at the free surface. \\

\noindent{\bf Supplementary movie caption}: \\ \\ \noindent{\it Movie 1:} Top panel shows the wavefield $\eta(r,t)$, scaled by the disc's oscillation amplitude $|Z|$. Bottom panel shows a colormap of the pressure with the hydrostatic contribution removed. The red line indicates the disc, which has radius $R$ and position $\zeta(t)$. The green arrow indicates the vertical applied force, which is proportional to $\cos\Omega t$. The parameters are $\text{Bo}=3$ and $\Omega=1$, which are the same as those used in figure~\ref{fig:WavePic}.
\\

\noindent{\bf Funding}: Support is acknowledged from NSF DMS-2510304 (AO) and NSF CBET-2338320 (DMH). 
\\

\noindent {\bf Declaration of interests}: The authors report no conflict of interest.
\\

\noindent {\bf Data availability statement}: The experimental data in figures~\ref{fig:ExpTheory} and~\ref{fig:ExpExtra} are available at \href{https://github.com/harrislab-brown/VerticalDiskDynamics}{https://github.com/harrislab-brown/VerticalDiskDynamics}.
\\

\noindent{\bf Author ORCIDs}: 

\noindent A. U. Oza, https://orcid.org/0000-0002-9079-9172

\noindent J.-W. Barotta, https://orcid.org/0000-0001-6769-5132

\noindent D. M. Harris, https://orcid.org/0000-0003-2615-9178

\appendix

\section{\markblue{Satisfying the radiation condition}}\label{App:RadCond}

We here show that the expression for $H(r)$ in Eq.~\eqref{h1Soln2} satisfies the radiation condition~\eqref{RadCond}. To that end, since $\text{K}_0(r)$ decays exponentially for large $r$ but $\text{J}_0(r)$ decays only algebraically, we have
\begin{align}
H(r)&\sim  -\mathrm{i}\Omega\text{ Bo}\lim_{\epsilon\rightarrow 0^+}\int_0^{\infty}\mathrm{d} k\,\frac{k}{(k-q_{\text{i}})(k-\overline{q}_{\text{i}})}\frac{\tilde{A}(k)}{k-q_{\text{r}}+\rmi\epsilon}\text{J}_0(kr)\quad\text{as}\quad r\rightarrow\infty.\label{HAsym1}
\end{align}
Following the argument presented by \citet{Miles1987}, we substitute the integral representation 
\begin{align}
\text{J}_0(kr)=\frac{1}{\pi}\int_0^{\pi/2}\rmd\theta\left(\rme^{\rmi kr\cos\theta}+\rme^{-\rmi kr\cos\theta}\right)
\end{align}
into Eq.~\eqref{HAsym1} and deform the path of integration for the $\exp(\rmi kr\cos\theta)$ and $\exp(-\rmi kr\cos\theta)$ terms to the positive and negative imaginary $k$-axis, respectively. Only the latter gives a nonzero contribution because the integrand in Eq.~\eqref{HAsym1} has a pole in the lower half-plane (at $k=q_{\text{r}}-\rmi\epsilon$), which yields
\begin{align}
H(r)&\sim -\frac{2\Omega\text{ Bo }q_{\text{r}}\tilde{A}(q_{\text{r}})}{\markblue{3q_{\text{r}}^2+\text{Bo}}}\int_0^{\pi/2}\rmd\theta\,\rme^{-\rmi q_{\text{r}}r\cos\theta}\quad\text{as}\quad r\rightarrow\infty.
\end{align}
By the stationary phase approximation and the Gaussian integral identity $\int_0^{\infty}\rmd\theta\,\rme^{\rmi x \theta^2}=\rme^{\rmi\pi/4}\sqrt{\pi/(4x)}$ for $x>0$, we obtain
\begin{align}
H(r)&\sim-\sqrt{\frac{2\pi q_{\text{r}}}{r}}\frac{\Omega\text{ Bo }\tilde{A}(q_{\text{r}})}{\markblue{3q_{\text{r}}^2+\text{Bo}}}\rme^{-\rmi\left(q_{\text{r}}r-\pi/4\right)}\quad\text{as}\quad r\rightarrow\infty,\label{HAsym2}
\end{align}
from which it is clear that Eq.~\eqref{RadCond} is satisfied.

\section{\markblue{Local behavior of the solution near the edge of the disc}}\label{App:Local}

\markblue{We here show that the behavior of the velocity potential $\Phi(r,0)$ and vertical fluid velocity $\partial_z\Phi(r,0)$ near the edge of the disc is described by Eq.~\eqref{LocalEqn}. We do so using two different approaches, for the sake of completeness: in Appendix~\ref{App:LocalIntEq}, we use the high-frequency behavior of the solution $A(k)$ of the integral equation~\eqref{IntEq}. In Appendix~\ref{App:LocalPDE}, we conduct a local analysis of the governing equations~\eqref{BathEqnND} near the edge of the disc, $r\approx 1$. We conclude in Appendix~\ref{App:LocalComparison} by comparing our results with prior results from the literature in which surface tension is neglected, $\text{Bo}=\infty$.}

\subsection{\markblue{Using the integral equation}}\label{App:LocalIntEq}

\markblue{In Appendix~\ref{App:HighFreq}, we show that the solution $A(k)$ to the integral equation~\eqref{IntEq} has the following high-frequency behavior:
\begin{align}
A(k)\sim A_\infty\frac{\text{J}_0(k)}{k^3}\quad\text{as}\quad k\rightarrow\infty,\label{AInf}
\end{align}
where $A_\infty=-\rmi\Omega H^{\prime}(1)$. The local behavior of $\Phi(r,0)$ for $r\approx 1$ is thus determined by 
\begin{align}
P(r)\equiv A_{\infty}\int_{k_0}^\infty \rmd k\,\frac{\text{J}_0(k)\text{J}_0(kr)}{k^2},\label{PDef}
\end{align}
while the behavior of $\partial_z\Phi(r,0)$ is determined by
\begin{align}
W(r)\equiv A_{\infty}\int_{k_0}^\infty\rmd k\,\frac{\text{J}_0(k)\text{J}_0(kr)}{k},\label{WDef}
\end{align}
where $k_0 > 0$ is fixed. To see this, recall the asymptotic formula for the Bessel function,
\begin{align}
\text{J}_n(x)\sim\sqrt{\frac{2}{\pi x}}\cos\left(x-\frac{\pi}{4}-\frac{n\pi}{2}\right)\quad\text{as}\quad x\rightarrow\infty,\label{BesselAsym}
\end{align}
and thus observe that the remainders $\Phi(r,0)-P(r)$ and $\partial_z\Phi(r,0)-W(r)$ are $C^2$ and $C^1$ at $r=1$, respectively. The cutoff $k_0$ is required simply because the integrands are not integrable at $k=0$.}

\markblue{To determine the behavior of $P(r)$ and $W(r)$ for $r\approx 1$, we use the following Sonine-Schafheitlin formulas:
\begin{align}
\int_0^\infty\rmd k\,\text{J}_0(k)\text{J}_1(kr)&=\begin{cases} 0 &\text{if } 0< r < 1, \\ \frac{1}{r}&\text{if }r > 1,\end{cases}\nonumber \\
\int_0^{\infty}\rmd k\,\frac{\text{J}_0(k)\text{J}_1(kr)}{k}&=\frac{2}{\pi r}\begin{cases} E(r)+(r^2-1)K(r)&\text{if }0< r < 1, \\ rE(1/r)&\text{if }r >1,\end{cases}
\end{align}
where $K(r)$ and $E(r)$ are the complete elliptic integrals of the first and second kind, respectively. Differentiating Eqs.~\eqref{PDef} and~\eqref{WDef} under the integral sign, with the proviso $r\neq 1$ for the latter because the corresponding integral converges only conditionally there, we obtain
\begin{align}
P^{\prime}(r)=\tilde{p}(r)-\frac{2A_\infty}{\pi r}\begin{cases} E(r)+(r^2-1)K(r)&\text{if }0< r < 1, \\ rE(1/r)&\text{if }r >1,\end{cases}\label{PPrime}
\end{align}
and
\begin{align}
W^{\prime}(r)=\tilde{w}(r)+\begin{cases} 0&\text{if }0 < r < 1, \\ -A_{\infty}/r&\text{if }r > 1,\end{cases}\label{WPrime}
\end{align}
where $\tilde{p}(r)$ and $\tilde{w}(r)$ are analytic at $r=1$. 
Using the asymptotic expansions~\citep[Eqs.~19.12.1 and~19.12.2]{NIST:DLMF}
\begin{align}
K(r)&\sim \frac{1}{2}\log\frac{8}{1-r}+O\left((1-r)\log|1-r|\right)\nonumber \\
\text{and}\quad E(r)&\sim 1+\frac{1-r}{2}\left(\log\frac{8}{1-r}-1\right)+O\left((1-r)^2\log|1-r|\right)\quad\text{as}\quad r\rightarrow 1^-,
\end{align}
we obtain
\begin{align}
P^{\prime}(r)&= \hat{p}_1(r)+\frac{A_\infty}{\pi}(r-1)\log|r-1|+O\left((r-1)^2\log|r-1|\right)\quad\text{as}\quad r\rightarrow 1,
\end{align}
and thus
\begin{align}
P(r)&=\hat{p}_0(r)+\frac{A_\infty}{2\pi}(r-1)^2\log|r-1| +O\left((r-1)^3\log|r-1|\right)\quad\text{as}\quad r\rightarrow 1,
\end{align}
where $\hat{p}_i(r)$ are analytic at $r=1$. Integrating Eq.~\eqref{WPrime} and using the fact that $W$ is continuous in $r$, because the integral converges absolutely and uniformly, we obtain
\begin{align}
W(r)&=\tilde{W}(r)+\begin{cases} 0 &\text{if }0 <r <1 \\ -A_\infty\log r&\text{if }r > 1,\end{cases}\nonumber \\
&=\tilde{W}(r)-A_\infty(r-1)_++O\left((r-1)_+^2\right)\quad\text{as}\quad r\rightarrow 1, 
\end{align}
where $\tilde{W}(r)$ is analytic at $r=1$. We thus conclude that
\begin{align}
\Phi(r,0)&=\hat{P}(r)+\frac{A_\infty}{2\pi}(r-1)^2\log(|r-1|) +O\left((r-1)^3\log|r-1|\right), 
\nonumber \\
\partial_z\Phi(r,0)&=\hat{W}(r)+\rmi\Omega Z-A_\infty(r-1)_++O\left((r-1)_+^2\right)\quad\text{as}\quad r\rightarrow 1,
\end{align}
where $\hat{P}(r)$ and $\hat{W}(r)$ are $C^2$ and $C^1$ at $r=1$, respectively. Note that, since $\partial_z\Phi(r,0)=\rmi\Omega Z$ on $r < 1$, $\hat{W}(r)\equiv 0$ for $r\in (0,1)$, and since $\hat{W}(r)\in C^1$, we have $\hat{W}(1)=\hat{W}^{\prime}(1)=0$. We thus have
\begin{align}
\partial_z\Phi(r,0)&=\rmi\Omega Z-A_\infty(r-1)_++o(r-1)\quad\text{as}\quad r\rightarrow 1,\label{dzPhiLocal}
\end{align}
}
\markblue{which establishes~\eqref{LocalEqn}.}

\subsubsection{\markblue{High-frequency behavior of $A(k)$}}\label{App:HighFreq}

\markblue{We now prove Eq.~\eqref{AInf}. Using Eq.~\eqref{BesselAsym}, we obtain the asymptotic formulas
\begin{align}
b(k)\sim\frac{\beta\text{J}_0(k)}{k^2}\quad\text{as}\quad k\rightarrow\infty
\end{align}
and
\begin{align}
F(q,k)&\sim\frac{q\text{ Bo}}{\text{Bo}+q^2}\left(\beta\text{J}_0(q)-q\text{J}_1(q)\right)\frac{\text{J}_0(k)}{k^2}\quad\text{as}\quad k\rightarrow\infty\quad\text{for }q\text{ fixed}.
\end{align}
In the limit $k\rightarrow\infty$, the integral equation~\eqref{IntEqReg} reads 
\begin{align}
\tilde{A}(k)&+\Omega^2\text{ Bo }\frac{\text{J}_0(k)}{k^2}\lim_{\epsilon\rightarrow 0^+}\int_0^{\infty}\rmd q\,q\frac{\beta\text{J}_0(q)-q\text{J}_1(q)}{(q-q_{\text{i}})(q-\overline{q}_{\text{i}})(q-q_{\text{r}}+\rmi\epsilon)}\tilde{A}(q)\nonumber \\
&\sim\rmi\Omega Z\beta\frac{\text{J}_0(k)}{k^2}\quad\text{as}\quad k\rightarrow\infty.
\end{align}
We thus conclude that
\begin{align}
\tilde{A}(k)\sim A_\infty \frac{\text{J}_0(k)}{k^2}\quad\text{as}\quad k\rightarrow\infty,
\end{align}
where
\begin{align}
A_\infty&=\rmi\Omega Z\beta-\Omega^2\text{ Bo }\lim_{\epsilon\rightarrow 0^+}\int_0^\infty\rmd q\,q\frac{\beta\text{J}_0(q)-q\text{J}_1(q)}{(q-q_\text{i})(q-\overline{q}_\text{i})(q-q_{\text{r}}+\rmi\epsilon)}\tilde{A}(q)\nonumber \\
&=-\rmi\Omega H^{\prime}(1)
\end{align}
from Eq.~\eqref{hp}.}

\subsection{\markblue{Local analysis of the PDEs}}\label{App:LocalPDE}

\markblue{We now give an alternative derivation of Eq.~\eqref{LocalEqn}, based on a local analysis of the governing equations~\eqref{BathEqnND} near the edge of the disc. Define the coordinates
\begin{align}
x=r-1\quad \text{and}\quad y=z,
\end{align}
so the edge of the disc is at the origin $(x,y)=(0,0)$. Since $\Delta\Phi\approx \partial_{xx}\Phi$ and $\Delta H\approx \partial_{xx}H$, the local problem is
\begin{subequations}
\begin{alignat}{2} \partial_{xx}\Phi+\partial_{yy}\Phi&=0,\quad &&x\in\mathbb{R},\quad y < 0,\label{LaplaceLocal} \\
\rmi\Omega\text{ Bo }\Phi&=-\text{Bo }H+\partial_{xx}H, \quad && x > 0,\quad y=0,\label{DynBCLocal} \\
\partial_y\Phi&=\rmi\Omega H,\quad && x > 0,\quad y=0,\label{KinBCLocal} \\
\partial_y\Phi&=\rmi\Omega Z,\quad && x<0,\quad y=0,\label{NoPenBCLocal} \\
H(0)&=Z.\label{PinBCLocal}
\end{alignat}\label{LocalPDE}
\end{subequations}
From Eqs.~\eqref{KinBCLocal}--\eqref{PinBCLocal}, we observe that $\partial_y\Phi(x,0)$ is continuous at $x=0$. Since the Neumann-to-Dirichlet operator for Laplace's equation is smoothing, $\Phi(x,0)$ is at least continuous at $x=0$, so it follows from Eq.~\eqref{DynBCLocal} that $H$ is differentiable at $x=0$. We can thus write
\begin{align}
\Phi(x,y)=\rmi\Omega Zy+\tilde{\Phi}(x,y)\quad\text{and}\quad  H(x)\approx Z+h_1x \quad\text{for}\quad (x,y)\approx (0,0),
\end{align}
where $\tilde{\Phi}$ solves the problem
\begin{subequations}
\begin{alignat}{2}
\partial_{xx}\tilde{\Phi}+\partial_{yy}\tilde{\Phi}&=0,\quad &&x\in\mathbb{R},\quad y < 0,\label{LaplaceLocal2} \\
\partial_y\tilde{\Phi}&=\rmi\Omega h_1x,\quad &&x > 0,\quad y=0,\label{KinBCLocal2} \\
\partial_y\tilde{\Phi}&=0,\quad &&x<0,\quad y=0.\label{NoPenBCLocal2}
\end{alignat}\label{LocalPDE2}
\end{subequations}
The local behavior of the solution to Eq.~\eqref{LocalPDE2} can be determined using standard techniques from complex analysis. Specifically, consider the analytic function $f(z)=\tilde{\Phi}(x,y)+\rmj \tilde{\Psi}(x,y)$, where $z=x+\rmj y$, and we use the imaginary unit $\rmj$ to avoid confusion with the time-harmonic factor $\rmi$. By the Cauchy-Riemann equations, Eqs.~\eqref{KinBCLocal2}-\eqref{NoPenBCLocal2} can be written as
\begin{subequations}
\begin{alignat}{2}
\partial_x\tilde{\Psi}&=-\rmi\Omega h_1x,\quad &&x > 0,\quad y=0, \\
\partial_x\tilde{\Psi}&=0,\quad &&x < 0,\quad y=0,
\end{alignat}\label{PsixBC}
\end{subequations}
which can be integrated to obtain
\begin{subequations}
\begin{alignat}{2}
\tilde{\Psi}&=\tilde{\Psi}_0-\frac{\rmi\Omega h_1}{2}x^2,\quad &&x>0,\quad y=0,\label{PsiKinBC} \\
\tilde{\Psi}&=\tilde{\Psi}_0,\quad &&x < 0,\quad y=0,\label{PsiNoPenBC}
\end{alignat}\label{PsiBC}
\end{subequations}
where $\tilde{\Psi}_0$ is a constant. To solve this problem, we propose the ansatz
\begin{align}
f(z)=p_0+p_1z+\frac{p_2}{2}z^2+Cz^2\log z.
\end{align}
To find the values of the constants $p_i$ and $C$, we express $f(z)$ in polar coordinates through  $z=\rho\rme^{\mathrm{j}\theta}$, where $\rho > 0$ and $\theta\in (-\pi,0)$.
Enforcing the condition~\eqref{PsiNoPenBC} (at $\theta=-\pi$), we obtain
\begin{align}
\text{Im}[p_0]=\tilde{\Psi}_0,\quad \text{Im}[p_1]=0,\quad \text{Im}[C]=0,\quad \text{Re}[C]=\frac{\text{Im}[p_2]}{2\pi},
\end{align}
where $\text{Im}$ denotes the imaginary part with respect to $\rmj$. 
Enforcing the condition~\eqref{PsiKinBC} (at $\theta=0$), we find that $\text{Im}[p_2]=-\rmi\Omega h_1$. We thus obtain a solution in terms of the free constants $\text{Re}[p_i]$:
\begin{align}
f(z)=\text{Re}[p_0]+\mathrm{j}\tilde{\Psi}_0+\text{Re}[p_1]z+\frac{1}{2}\left(\text{Re}[p_2]-\mathrm{j}\rmi\Omega h_1\right)z^2-\frac{\rmi\Omega h_1}{2\pi}z^2\log z.
\end{align}
At the interface ($y=0$), we have
\begin{align}
\tilde{\Phi}(x,0)=Q(x)-\frac{\rmi\Omega h_1}{2\pi}x^2\log |x|,\label{PhiTildeSoln}
\end{align}
where $Q(x)$ is a quadratic polynomial in $x$. Note that the homogeneous problem corresponding to~\eqref{PsixBC} has eigenfunctions $\rho^n\cos n\theta$ for $n\in\mathbb{Z}$. The solutions for $n\leq -1$, along with $\log\rho$, diverge as $\rho\rightarrow 0$, so they are excluded by continuity of $\Phi$. The remaining solutions ($n\geq 0$) are analytic, so the general solution to~\eqref{LocalPDE2} is the sum of~\eqref{PhiTildeSoln} and an analytic function. We also have
\begin{align}
\partial_y\tilde{\Phi}(x,0)=-\left.\text{Im}\left[f^{\prime}(z)\right]\right|_{y=0}=\rmi\Omega h_1x_+.\label{dyPhiTilde}
\end{align}
Since $h_1=H^{\prime}(1)$ by definition, we conclude that~\eqref{PhiTildeSoln} and~\eqref{dyPhiTilde} are consistent with~\eqref{LocalEqn}.}

\subsection{\markblue{Comparison with  the gravity-only case ($\mathrm{Bo}=\infty$)}}\label{App:LocalComparison}

\markblue{We now consider the edge behavior in the gravity-only case, for which surface tension is neglected ($\text{Bo}=\infty$), and compare that to Eq.~\eqref{LocalEqn}. While the gravity-only case has been treated extensively in the literature, we briefly summarize the derivation for the sake of completeness, and to connect the steps with Appendix~\ref{App:LocalPDE}. 
The relevant boundary value problem is
\begin{subequations}
\begin{alignat}{2}
\Delta\Phi+\partial_{zz}\Phi&=0,\quad &&r > 0,\quad z < 0, \\
\partial_z\Phi&=\rmi\Omega Z,\quad &&r < 1,\quad z=0, \\
\partial_z\Phi&=\Omega^2\Phi,\quad &&r > 1,\quad z =0.
\end{alignat}
\end{subequations}
Introducing the coordinates $(x,y)$ and approximating $\Delta\Phi\approx \partial_{xx}\Phi$ as in Appendix~\ref{App:LocalPDE}, we write $\Phi(x,y)=\rmi\Omega Zy+\rmi Z/\Omega+\tilde{\Phi}(x,y)$ and obtain the problem
\begin{subequations}
\begin{alignat}{2}
\partial_{xx}\tilde{\Phi}+\partial_{yy}\tilde{\Phi}&=0,\quad &&x\in\mathbb{R},\quad y<0, \\
\partial_y\tilde{\Phi}&=0,\quad &&x < 0,\quad y=0, \label{NoPenBCGrav}\\
\partial_y\tilde{\Phi}&=\Omega^2\tilde{\Phi},\quad &&x > 0,\quad y=0.\label{KinBCGrav}
\end{alignat}\label{PhiTildePDEGrav}
\end{subequations}
As in Appendix~\ref{App:LocalPDE}, we solve this problem by recasting it in terms of an analytic function $f(z)=\tilde{\Phi}(x,y)+\rmj \tilde{\Psi}(x,y)$, where $z=x+\rmj y$. We here propose the ansatz 
\begin{align}
f(z) = p_0+p_1z+Cz\log z. 
\end{align}
By enforcing the boundary conditions~\eqref{NoPenBCGrav} and~\eqref{KinBCGrav}, we find that $\text{Im}[C]=0$, $\text{Im}[p_1]=-\Omega^2\text{ Re}[p_0]$ and $\text{Re}[C]=\text{Im}[p_1]/\pi$, where again $\text{Im}$ denotes the imaginary part with respect to the complex unit $\rmj$. We thus obtain a solution in terms of the free constants $p_0$ and $\text{Re}[p_1]$,
\begin{align}
f(z) &= p_0+\left(\text{Re}[p_1]-\rmj\Omega^2\text{ Re}[p_0]\right)z-\frac{\Omega^2\text{ Re}[p_0]}{\pi}z\log z,
\end{align}
which is consistent with that obtained by~\citet{Friedrichs1948}. The local behavior of the velocity potential and vertical velocity near the edge of the disc is thus
\begin{align}
\tilde{\Phi}(x,y)&=\text{Re}[p_0]\left[1+\Omega^2y-\frac{\Omega^2}{\pi}\left(x\log \rho-y\theta\right)\right]+\text{Re}[p_1]x,\nonumber \\
\partial_y\tilde{\Phi}(x,y)&=\Omega^2\text{ Re}[p_0]\left(1+\frac{\theta}{\pi}\right),\label{PhiTildeSolnGrav}
\end{align}
where $x=\rho\cos\theta$ and $y=\rho\sin\theta$ as in Appendix~\ref{App:LocalPDE}.}

\markblue{We make two observations about the solution~\eqref{PhiTildeSolnGrav}. First, the velocity potential contains a term proportional to $x\log|x|$ as $x\rightarrow 0$ for $y=0$, as opposed to the smoother $x^2\log|x|$ behavior obtained in Eq.~\eqref{PhiTildeSoln} when surface tension is considered. Second, the vertical fluid velocity at the interface ($y=0$) jumps by $\Omega^2\text{ Re}[p_0]$ across $x=0$. That is, the vertical fluid velocity is discontinuous across the edge of the disc, unlike the continuous (but not differentiable) behavior predicted by~\eqref{dyPhiTilde} when surface tension is considered.
}

\section{\markblue{Convergence of the numerical algorithm}}\label{App:Convergence}

\markblue{We here provide evidence for the convergence of the numerical algorithm for solving the integral equation~\eqref{IntEqReg}, as described in \S~\ref{SSec:Numerical}. To that end, we fix the parameter values $\Omega=1$ and $\text{Bo}=3$, and vary the total number of points $N$ from 50 to 5000. The numerical parameters are set as $a=10q_{\text{r}}$, $k_{\text{max}}=200q_{\text{r}}$, $N_1=N/25$ and $N_2=9N/25$. For each $N$, the error in $\Phi$, which we call $E_\Phi$, is assessed by computing $2\pi\int_0^1\rmd r\,r\Phi(r,0)$ using Eq.~\eqref{PhiInt}, and comparing its value to that obtained from using a high resolution of $N=10^4$, which is twice that adopted in the paper. Similarly, the error in $H$, which we call $E_H$, is assessed by computing $H^{\prime}(1)$ using Eq.~\eqref{hp}. \\ \indent Figure~\ref{fig:Convergence} shows the dependence of the errors $E_\Phi$ and $E_H$ on the number of points $N$. We observe that both errors decrease with $N$, assuming values of order $10^{-6}$ for the value of $N=5000$ adopted for the calculations in the paper. We thus conclude that the computations presented herein are numerically well-resolved.}

\begin{figure}
    \centering
    \includegraphics[width = 1\textwidth]{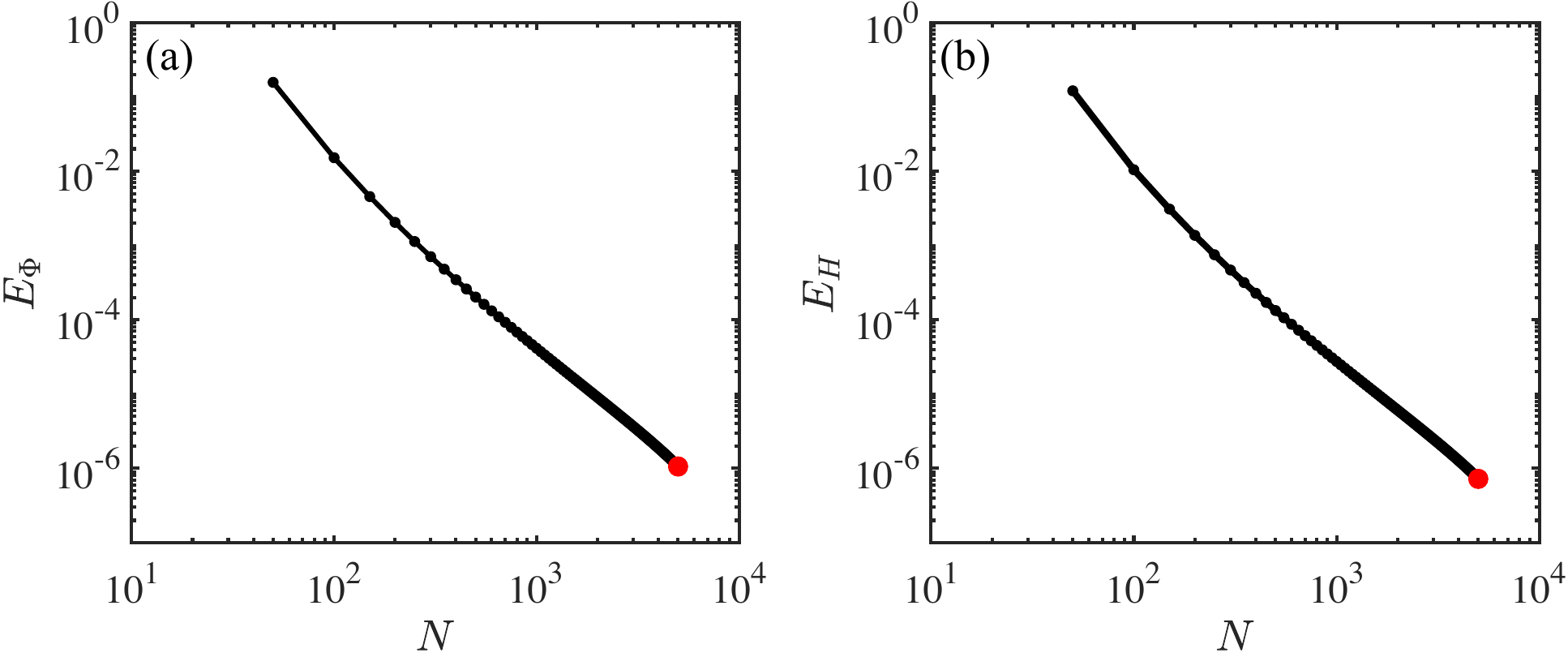}
    \caption{\markblue{Convergence of the numerical algorithm described in \S~\ref{SSec:Numerical}. The plots show the dependence of the errors $E_\Phi$ (panel~\textit{(a)}) and $E_H$ (panel~\textit{(b)}) on the total number of points $N$. The value $N=5000$ (red dot) is adopted for the numerical calculations presented herein.}}
    \label{fig:Convergence}
\end{figure}

\section{Additional experimental data sets}\label{App:Extra}

To complement the results in figure~\ref{fig:ExpTheory}, we conducted experiments for a total of six different discs. The discs varied in both radius and mass, yielding different $(\text{Bo},M)$ parameter combinations. Overall, we see favorable agreement between experiment and theory for all six discs tested, as shown in figure \ref{fig:ExpExtra}. 

\begin{figure}
    \centering
    \includegraphics[width = .9\textwidth]{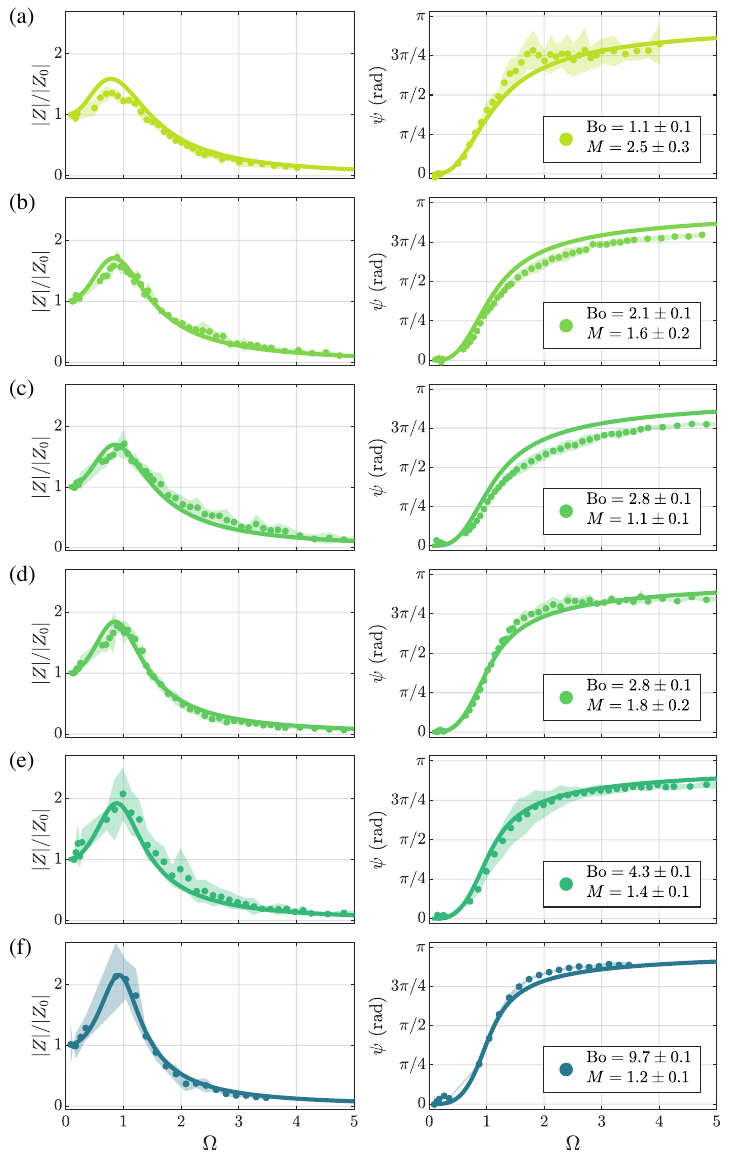}
    \caption{\textit{(a-f)} Transmissibility and phase lag for six different discs, with parameters listed in each plot legend. The mean values of the experimental data across six independent trials are shown as points, with one standard deviation shown as the shaded region between the means of the trials. The corresponding theoretical predictions are shown as the solid curves. The data in panels (a), (c), and (f) are also shown in figure~\ref{fig:ExpTheory}.}
    \label{fig:ExpExtra}
\end{figure}

\section{Asymptotics of physical quantities in the low frequency limit $\Omega\rightarrow 0$}\label{App:Asymptotics}

To obtain expressions for the added mass $M_{\text{P}}^0$ and capillary spring constant correction $\kappa$ in the low-frequency limit $\Omega\rightarrow 0$, we need to evaluate the integrals in Eqs.~\eqref{LowOmM} and~\eqref{LowOmk}, respectively. To that end, we substitute the expression for $b(k)$ from Eq.~\eqref{IntEq} into the integrals and use the facts that
\begin{align}
b_0\equiv \int_0^{\infty}\rmd k\,\frac{\text{J}_1^2(k)}{k^2}&=\frac{4}{3\pi},\nonumber \\
b_1\equiv \int_0^{\infty}\rmd k\,\frac{\text{J}_1^2(k)}{\text{Bo}+k^2}&=-\frac{\pi}{2\sqrt{\text{Bo}}}\text{I}_1^2\left(\sqrt{\text{Bo}}\right)+\frac{4}{3\pi}\,{}_2F_3\left(1,1;\frac{1}{2},\frac{3}{2},\frac{5}{2};\text{Bo}\right),\nonumber \\
b_2\equiv \int_0^{\infty}\rmd k\,\frac{\text{J}_0(k)\text{J}_1(k)}{k(\text{Bo}+k^2)}&=\frac{\pi}{2\text{ Bo}}\text{I}_0\left(\sqrt{\text{Bo}}\right)\text{I}_1\left(\sqrt{\text{Bo}}\right)-\frac{16}{9\pi}\,{}_2F_3\left(1,2;\frac{3}{2},\frac{5}{2},\frac{5}{2};\text{Bo}\right),\nonumber \\ 
b_3\equiv \int_0^{\infty}\rmd k\,\frac{k\text{J}_0(k)\text{J}_1(k)}{\left(\text{Bo}+k^2\right)^2}&=\frac{\pi}{4\sqrt{\text{Bo}}}\left[\text{I}_0^2\left(\sqrt{\text{Bo}}\right)+\text{I}_1^2\left(\sqrt{\text{Bo}}\right)-\frac{1}{\sqrt{\text{Bo}}}\text{I}_0\left(\sqrt{\text{Bo}}\right)\text{I}_1\left(\sqrt{\text{Bo}}\right)\right]\nonumber \\
&\phantom{=}-\frac{16}{9\pi}{}_2F_3\left(2,2;\frac{3}{2},\frac{5}{2},\frac{5}{2};\text{Bo}\right),\nonumber \\
b_4\equiv \int_0^{\infty}\rmd k\,\frac{\text{J}_0^2(k)}{\left(\text{Bo}+k^2\right)^2}&=\frac{\pi}{4\text{ Bo}^{3/2}}\text{I}_0\left(\sqrt{\text{Bo}}\right)\left[\text{I}_0\left(\sqrt{\text{Bo}}\right)-2\sqrt{\text{Bo}}\,\text{I}_1\left(\sqrt{\text{Bo}}\right)\right]\nonumber \\
&\phantom{=}+\frac{32}{27\pi}{}_2F_3\left(2,2;\frac{5}{2},\frac{5}{2},\frac{5}{2};\text{Bo}\right),\nonumber \\
b_5\equiv \int_0^{\infty}\rmd k\,\frac{k^2\text{J}_1^2(k)}{(\text{Bo}+k^2)^2}&=\frac{\pi}{2}\text{I}_1\left(\sqrt{\text{Bo}}\right)\left[-\text{I}_0\left(\sqrt{\text{Bo}}\right)+\frac{1}{2\sqrt{\text{Bo}}}\text{I}_1\left(\sqrt{\text{Bo}}\right)\right]\nonumber \\
&\phantom{=}+\frac{4}{3\pi}{}_2F_3\left(1,2;\frac{1}{2},\frac{3}{2},\frac{5}{2};\text{Bo}\right),\label{biInts}
\end{align}
where ${}_pF_q$ is the generalized hypergeometric function. We thus obtain
\begin{align}
M_{\text{P}}^0=2\pi \left(b_0-b_1+\beta b_2\right),
\end{align}
which is equal to Eq.~\eqref{MP0} by virtue of Eq.~\eqref{biInts}. Similarly, we obtain
\begin{align}
\text{ Bo}^{-1}\beta\kappa&=b_1-\beta b_2+2\beta b_3-\beta^2b_4-b_5\nonumber \\
&=\frac{\pi}{2\sqrt{\text{Bo}}}\left[\text{I}_0^2\left(\sqrt{\text{Bo}}\right)\beta\left(1-\frac{\beta}{2\text{ Bo}}\right)+\text{I}_1^2\left(\sqrt{\text{Bo}}\right)\left(\beta-\frac{3}{2}\right)\right.\nonumber \\
&\phantom{=}\left.+\text{I}_0\left(\sqrt{\text{Bo}}\right)\text{I}_1\left(\sqrt{\text{Bo}}\right)\left(\frac{\beta(\beta-2)}{\sqrt{\text{Bo}}}+\sqrt{\text{Bo}}\right)\right]+\frac{4}{3\pi}\,{}_2F_3\left(1,1;\frac{1}{2},\frac{3}{2},\frac{5}{2};\text{Bo}\right)\nonumber \\
&\phantom{=}+\frac{16\beta}{9\pi}\,{}_2F_3\left(1,2;\frac{3}{2},\frac{5}{2},\frac{5}{2};\text{Bo}\right)-\frac{32\beta}{9\pi}\,{}_2F_3\left(2,2;\frac{3}{2},\frac{5}{2},\frac{5}{2};\text{Bo}\right)\nonumber \\
&\phantom{=}-\frac{4}{3\pi}{}_2F_3\left(1,2;\frac{1}{2},\frac{3}{2},\frac{5}{2};\text{Bo}\right)-\frac{32\beta^2}{27\pi}\,{}_2F_3\left(2,2;\frac{5}{2},\frac{5}{2},\frac{5}{2};\text{Bo}\right).\label{kappaST}
\end{align}

To obtain expressions for $M_{\text{P}}^0$ and $\kappa$ in the limit $\text{Bo}\rightarrow 0$, we require the following expansions of $b_i$, obtained using Mathematica: 
\begin{align}
b_1,\, b_5&\sim \frac{4}{3\pi},\quad b_2\sim \frac{\pi}{4\sqrt{\text{Bo}}}-\frac{16}{9\pi},\quad
b_3\sim \frac{\pi}{8\sqrt{\text{Bo}}}-\frac{16}{9\pi},\nonumber \\
\text{and}\quad b_4&\sim \frac{\pi}{4\text{ Bo}^{3/2}}-\frac{\pi}{8\sqrt{\text{Bo}}}+\frac{32}{27\pi}\quad\text{as}\quad \text{Bo}\rightarrow 0.
\end{align}
We also have
\begin{align}
\beta\sim -\frac{2}{2\gamma+\log(\text{Bo}/4)}\quad\text{as}\quad \text{Bo}\rightarrow 0,
\end{align}
and thus obtain Eqs.~\eqref{MP0Bo0} and~\eqref{kappaBo0}.

To obtain approximations for $M_{\text{P}}^0$ and $\kappa$ in the limit $\text{Bo}\rightarrow\infty$, we determine the leading order behavior of the coefficients $b_i$. Starting with $b_1$, we have
\begin{align}
b_1=\int_0^{\markblue{k_\infty}}\rmd k\,\frac{\text{J}_1^2(k)}{\text{Bo}+k^2}+\int_{\markblue{k_\infty}}^\infty\rmd k\,\frac{\text{J}_1^2(k)}{\text{Bo}+k^2},
\end{align}
{\markblue{where $k_{\infty}\gg 1$ is a cutoff value.}} The first integral is clearly $O(1/\text{Bo})$. For the second, \markblue{we recall the asymptotic formula}~\eqref{BesselAsym}, from which it follows that 
\begin{align}
\int_{\markblue{k_\infty}}^\infty\rmd k\,\frac{\text{J}_1^2(k)}{\text{Bo}+k^2}&\markblue{\approx\frac{2}{\pi}\int_{\markblue{k_\infty}}^{\infty}\rmd k\,\frac{\cos^2\left(k-\frac{3\pi}{4}\right)}{k\left(\text{Bo}+k^2\right)}=\frac{1}{\pi}\int_{k_\infty}^\infty\rmd k\,\frac{1-\sin 2k}{k\left(\text{Bo}+k^2\right)}}\nonumber \\
&\markblue{=\frac{1}{\pi}\int_{k_\infty}^\infty\rmd k\,\frac{1}{k\left(\text{Bo}+k^2\right)}+O\left(\frac{1}{\text{Bo}}\right)=\frac{\log\left(k_\infty^2+\text{Bo}\right)-2\log k_\infty}{2\pi\text{ Bo}}+O\left(\frac{1}{\text{Bo}}\right)}\nonumber \\
& \markblue{=\frac{\log\text{Bo}}{2\pi\text{ Bo}}+O\left(\frac{1}{\text{Bo}}\right)}\quad\text{as}\quad\text{Bo}\rightarrow\infty,
\end{align}
\markblue{so}
\begin{align}
    b_1=O\left(\frac{\log\text{Bo}}{\text{Bo}}\right)\quad\text{as}\quad \text{Bo}\rightarrow\infty.
\end{align}
By a similar argument, we obtain
\begin{align}
b_4= O\left(\frac{\log\text{Bo}}{\text{Bo}^2}\right),\quad b_5=O\left(\frac{1}{\text{Bo}}\right)\quad\text{as}\quad \text{Bo}\rightarrow\infty.
\end{align}
For $b_3$, we rewrite the integral using the identity $\text{J}_0^{\prime}(x)=-\text{J}_1(x)$ and integrate by parts:
\begin{align}
b_3=-\frac{1}{2}\int_0^{\infty}\rmd k\frac{k}{\left(\text{Bo}+k^2\right)^2}\diff{}{k}\left[\text{J}_0^2(k)\right]=\frac{1}{2}\int_0^{\infty}\rmd k\,\frac{\text{Bo}-3k^2}{\left(\text{Bo}+k^2\right)^3}\text{J}_0^2(k).
\end{align}
We thus obtain
\begin{align}
b_3= O\left(\frac{\log\text{Bo}}{\text{Bo}^2}\right)\quad\text{as}\quad \text{Bo}\rightarrow\infty.
\end{align}
The leading-order term for $b_2$ may be obtained simply as
\begin{align}
b_2\sim\frac{1}{\text{Bo}}\int_0^{\infty}\rmd k\,\frac{\text{J}_0(k)\text{J}_1(k)}{k}=\frac{2}{\pi\text{ Bo}}\quad\text{as}\quad \text{Bo}\rightarrow\infty.
\end{align}
We also have
\begin{align}
\beta\sim\sqrt{\text{Bo}}+\frac{1}{2}\quad\text{as}\quad \text{Bo}\rightarrow\infty.
\end{align}
We thus obtain the expressions for $M_\text{P}^0$ and $\kappa$ in the limit $\text{Bo}\rightarrow\infty$ as given in Eqs.~\eqref{MP0BoInf} and~\eqref{kappaBoInf}, respectively.

\bibliographystyle{jfm}
\bibliography{SurferBib}

@article{Islam2019,
    author = {Islam, Najnin and Kundu, Souvik and Gayen, Rupanwita},
    title = {Scattering and radiation of water waves by a submerged rigid disc in a two-layer fluid},
    journal = {Proceedings of the Royal Society A: Mathematical, Physical and Engineering Sciences},
    volume = {475},
    number = {2232},
    pages = {20190331},
    year = {2019},
    month = {12},
    issn = {1364-5021},
    doi = {10.1098/rspa.2019.0331},
    url = {https://doi.org/10.1098/rspa.2019.0331}
}

@article{Porter2015,
title = {Linearised water wave problems involving submerged horizontal plates},
journal = {Applied Ocean Research},
volume = {50},
pages = {91-109},
year = {2015},
issn = {0141-1187},
doi = {https://doi.org/10.1016/j.apor.2014.07.013},
url = {https://www.sciencedirect.com/science/article/pii/S0141118714000716},
author = {R. Porter}
}

@article{Yu1993,
  author    = {Yu, Xiping and Chwang, Allen T.},
  title     = {Analysis of Wave Scattering by Submerged Circular Disk},
  journal   = {Journal of Engineering Mechanics},
  volume    = {119},
  number    = {9},
  pages     = {1804--1817},
  year      = {1993},
  publisher = {American Society of Civil Engineers},
  doi       = {10.1061/(ASCE)0733-9399(1993)119:9(1804)},
  url       = {https://doi.org/10.1061/(ASCE)0733-9399(1993)119:9(1804)}
}

@article{Chwang1994,
  author    = {Chwang, Allen T. and Wu, Jianhua},
  title     = {Wave Scattering by Submerged Porous Disk},
  journal   = {Journal of Engineering Mechanics},
  volume    = {120},
  number    = {12},
  pages     = {2575--2587},
  year      = {1994},
  publisher = {American Society of Civil Engineers},
  doi       = {10.1061/(ASCE)0733-9399(1994)120:12(2575)},
  url       = {https://doi.org/10.1061/(ASCE)0733-9399(1994)120:12(2575)}
}

@article{Martin1997,
  author    = {Martin, P. A. and Farina, L.},
  title     = {Radiation of water waves by a heaving submerged horizontal disc},
  journal   = {Journal of Fluid Mechanics},
  volume    = {337},
  pages     = {365--379},
  year      = {1997},
  publisher = {Cambridge University Press},
  doi       = {10.1017/S0022112097004989},
  url       = {https://doi.org/10.1017/S0022112097004989}
}

@article{Das2025,
    author = {Das, Arijit and De, Soumen},
    title = {Water wave interaction with a submerged porous disc in a two-layer fluid},
    journal = {Physics of Fluids},
    volume = {37},
    number = {2},
    pages = {027158},
    year = {2025},
    month = {02},
    issn = {1070-6631},
    doi = {10.1063/5.0255719},
    url = {https://doi.org/10.1063/5.0255719}
}

@article{Das2023,
title = {Wave interaction with an elliptic disc submerged in a two-layer fluid},
journal = {Applied Mathematical Modelling},
volume = {117},
pages = {786-801},
year = {2023},
issn = {0307-904X},
doi = {https://doi.org/10.1016/j.apm.2023.01.016},
url = {https://www.sciencedirect.com/science/article/pii/S0307904X23000173},
author = {Arijit Das and Soumen De and B.N. Mandal}
}

@article{Friedrichs1948,
author = {Friedrichs, K. O. and Lewy, H.},
title = {The dock problem},
journal = {Communications on Pure and Applied Mathematics},
volume = {1},
number = {2},
pages = {135-148},
doi = {https://doi.org/10.1002/cpa.3160010203},
url = {https://onlinelibrary.wiley.com/doi/abs/10.1002/cpa.3160010203},
year = {1948}
}

@article{Yeh2008,
title = {The Exact Solutions of the Problems in Forced Capillary-Gravity Waves Generated by a Plane Wavemaker under {H}ocking's Edge Condition},
author = {Nai-Sher Yeh},
journal = {Taiwanese Journal of Mathematics},
volume = {12},
number = {1},
year = {2008}, 
pages = {1-24}
}

@article{Mandal1997,
title = {A note on the plane vertical wavemaker in the presence of surface tension},
author = {B. N. Mandal and S. Bandyopadhyay},
journal = {Quart. Appl. Math.},
volume = {49},
year = {1991},
pages = {627-633},
doi = {https://doi.org/10.1090/qam/1134744}
}

@article{Hocking1988b, 
title={Capillary-gravity waves with boundaries: Three-dimensional effects}, 
volume={10},
issue = {4}, 
DOI={},  
journal={Wave Motion}, 
author={L. M. Hocking}, 
year={1988}, 
pages={301--311}
}

@article{Young1987, 
title={A plate oscillating across a liquid interface: effects of contact-angle hysteresis}, 
volume={174}, 
DOI={},  
journal={Journal of Fluid Mechanics}, 
author={G. W. Young and S. H. Davis}, 
year={1987}, 
pages={327--356}
}

@article{Hocking1991, 
title={Capillary-gravity waves produced by a wavemaker}, 
volume={224}, 
DOI={},  
journal={Journal of Fluid Mechanics}, 
author={L. M. Hocking}, 
year={1991}, 
pages={217--226}
}

@article{Hocking1988a, 
title={Capillary-gravity waves produced by a heaving body}, 
volume={186}, 
DOI={},  
journal={Journal of Fluid Mechanics}, 
author={L. M. Hocking}, 
year={1988}, 
pages={337--349}
}

@article{Hocking1990, 
title={Scattering of a capillary–gravity wave by a vertical cylinder}, 
volume={2}, 
DOI={}, 
number={2}, 
journal={Physics of Fluids}, 
author={D. Mahdmina and L. M. Hocking}, 
year={1990}, 
pages={202--208}
}

@article{Harris2025Review,
  title = {Propulsion and interaction of wave-propelled interfacial particles},
  author = {Harris, Daniel M. and Barotta, Jack-William},
  journal = {Phys. Rev. Fluids},
  volume = {10},
  issue = {10},
  pages = {100503},
  numpages = {35},
  year = {2025},
  month = {Oct},
  publisher = {American Physical Society},
  doi = {10.1103/353x-p2dx},
  url = {https://link.aps.org/doi/10.1103/353x-p2dx}
}

@article{MacCamy1961a,
    author = {MacCamy, R. C.},
    title = {On the Heaving Motion of Cylinders of Shallow Draft},
    journal = {Journal of Ship Research},
    volume = {5},
    number = {04},
    pages = {34-43},
    year = {1961},
    issn = {0022-4502},
    doi = {10.5957/jsr.1961.5.4.34},
    url = {https://doi.org/10.5957/jsr.1961.5.4.34}
}

@article{Kim1963, 
title={The pitching motion of a circular disk}, 
volume={17}, 
DOI={10.1017/S0022112063001543}, 
number={4}, 
journal={Journal of Fluid Mechanics}, 
author={Kim, W. D.}, 
year={1963}, 
pages={607–629}
}

@article{Evans1968b, 
title={The effect of surface tension on the waves produced by a heaving circular cylinder}, 
volume={64}, 
DOI={10.1017/S030500410004353X}, 
number={3}, 
journal={Mathematical Proceedings of the Cambridge Philosophical Society}, 
author={Evans, D. V.}, 
year={1968}, 
pages={833–847}
}

@article{Kim1965, 
title={On the harmonic oscillations of a rigid body on a free surface}, 
volume={21}, 
DOI={10.1017/S0022112065000253}, 
number={3}, 
journal={Journal of Fluid Mechanics}, 
author={Kim, W. D.}, 
year={1965}, 
pages={427–451}
}

@article{Wehausen1971,
   author = "Wehausen, J V",
   title = "The Motion of Floating Bodies", 
   journal= "Annual Review of Fluid Mechanics",
   year = "1971",
   volume = "3",
   pages = "237-268",
   doi = "https://doi.org/10.1146/annurev.fl.03.010171.001321",
   url = "https://www.annualreviews.org/content/journals/10.1146/annurev.fl.03.010171.001321",
   publisher = "Annual Reviews",
   issn = "1545-4479",
   type = "Journal Article",
  }

@article{Hulme1982, 
title={The wave forces acting on a floating hemisphere undergoing forced periodic oscillations}, volume={121}, 
DOI={10.1017/S0022112082001980}, 
journal={Journal of Fluid Mechanics}, 
author={Hulme, A.}, 
year={1982}, 
pages={443–463}
}

@article{Havelock1955,
author = {Havelock, Thomas Henry},
title = {Waves due to a floating sphere making periodic heaving oscillations},
journal = {Proceedings of the Royal Society of London. Series A. Mathematical and Physical Sciences},
volume = {231},
number = {1184},
pages = {1-7},
year = {1955},
doi = {10.1098/rspa.1955.0152},
URL = {https://royalsocietypublishing.org/doi/abs/10.1098/rspa.1955.0152}
}

@article{Miles1987, 
title={On surface-wave forcing by a circular disk}, 
volume={175}, 
DOI={10.1017/S0022112087000302}, 
journal={Journal of Fluid Mechanics}, 
author={Miles, John W.}, 
year={1987}, 
pages={97–108}
}

@article{Sungar2024,
      title = {Synchronization and self-assembly of free capillary spinners},
  author = {Sungar, Nilgun and Sharpe, John and Ijzerman, Loic and Barotta, Jack-William},
  journal = {Phys. Rev. E},
  volume = {111},
  issue = {3},
  pages = {035104},
  numpages = {10},
  year = {2025},
  month = {Mar},
  publisher = {American Physical Society},
  doi = {10.1103/PhysRevE.111.035104},
  url = {https://link.aps.org/doi/10.1103/PhysRevE.111.035104}
}

@article{Yuan2012,
    title = {Bio-inspired micro/mini propulsion at air-water interface: A review},
  author = {Yuan, Junqi and Cho, Sung Kwon},
  journal = {Journal of Mechanical Science and Technology},
  volume = {26},
  number = {12},
  pages = {3761--3768},
  doi = {10.1007/s12206-012-1002-6},
  url = {https://doi.org/10.1007/s12206-012-1002-6},
  year = {2012}
}

@article{Rhee2022,
doi = {10.1088/1748-3190/ac78b6},
url = {https://dx.doi.org/10.1088/1748-3190/ac78b6},
year = {2022},
publisher = {IOP Publishing},
volume = {17},
number = {5},
pages = {055001},
author = {Eugene Rhee and Robert Hunt and Stuart J Thomson and Daniel M Harris},
title = {Surfer{B}ot: a wave-propelled aquatic vibrobot},
journal = {Bioinspiration \& Biomimetics}
}

@article{Barotta2024,
  title = {Synchronization of wave-propelled capillary spinners},
  author = {Barotta, Jack-William and Pucci, Giuseppe and Silver, Eli and Hooshanginejad, Alireza and Harris, Daniel M.},
  journal = {Phys. Rev. E},
  volume = {111},
  issue = {3},
  pages = {035105},
  numpages = {14},
  year = {2025},
  doi = {10.1103/PhysRevE.111.035105},
  url = {https://link.aps.org/doi/10.1103/PhysRevE.111.035105}
}

@article{Barotta2023,
    title = {Bidirectional wave-propelled capillary spinners},
  author = {Barotta, Jack-William and Thomson, Stuart J and Alventosa, Luke F. L and Lewis, Maya and Harris, Daniel M.},
  journal = {Communications Physics},
  volume = {6},
  issue = {1},
  pages = {87},
  year = {2023},
  doi = {10.1038/s42005-023-01206-z},
  url = {https://doi.org/10.1038/s42005-023-01206-z}
}

@article{Benham2024,
title={On wave-driven propulsion}, 
volume={987}, 
DOI={10.1017/jfm.2024.352}, 
journal={Journal of Fluid Mechanics}, 
author={Benham, Graham P. and Devauchelle, Olivier and Thomson, Stuart J.}, 
year={2024}, 
pages={A44}
}

@article{OzaSurfers,
    title = {Theoretical modeling of capillary surfer interactions on a vibrating fluid bath},
  author = {Oza, Anand U. and Pucci, Giuseppe and Ho, Ian and Harris, Daniel M.},
  journal = {Phys. Rev. Fluids},
  volume = {8},
  issue = {11},
  pages = {114001},
  numpages = {33},
  year = {2023},
  publisher = {American Physical Society},
  doi = {10.1103/PhysRevFluids.8.114001},
  url = {https://link.aps.org/doi/10.1103/PhysRevFluids.8.114001}
}

@misc{NIST:DLMF,
author = {{DLMF}},
year = {2026},
title = "{\it NIST Digital Library of Mathematical Functions}",
howpublished = "http://dlmf.nist.gov/, Version 1.2.7, Released on 2026-06-15",
url = "http://dlmf.nist.gov/",
note = "{F}.~W.~J. Olver, A.~B. {Olde Daalhuis}, D.~W. Lozier, B.~I. Schneider, R.~F. Boisvert, C.~W. Clark, B.~R. Miller, B.~V. Saunders, H.~S. Cohl, and M.~A. McClain, eds."}

@book{Lamb1932,
author = {H. Lamb},
publisher = {Cambridge University Press},
title = {Hydrodynamics},
year = {1932}
}

@article{Dias2008,
  title={Theory of weakly damped free-surface flows: A new formulation based on potential flow solutions},
  author={F. Dias and A. Dyachenko and V. E. Zakharov},
  journal={Phys. Lett. A},
  volume={372},
  pages={1297--1302},
  year={2008}
}

@article{faraday1831forms,
  title={On the forms and states assumed by fluids in contact with elastic vibrating surfaces},
  author={Faraday, M},
  journal={Philos. Trans. R. Soc. London},
  volume={121},
  pages={319},
  year={1831}
}

@article{Ursell1949,
  title={On the heaving motion of a circular cylinder on the surface of a fluid},
  author={F. Ursell},
  journal={Quart. J. Mech. Appl. Math.},
  volume={2},
  pages = {218-231},
  year={1949},
}

@inproceedings{longuet1964radiation,
  title={Radiation stresses in water waves; a physical discussion, with applications},
  author={Longuet-Higgins, Michael S. and Stewart, R.W.},
  booktitle={Deep Sea Research and Oceanographic Abstracts},
  volume={11},
  number={4},
  pages={529--562},
  year={1964},
  organization={Elsevier}
}

@article{de2018capillary,
  title={Capillary interactions between dynamically forced particles adsorbed at a planar interface and on a bubble},
  author={De Corato, M and Garbin, V},
  journal={Journal of Fluid Mechanics},
  volume={847},
  pages={71--92},
  year={2018},
  publisher={Cambridge University Press}
}

@article{RhodesRobinson1971, 
title={On the forced surface waves due to a vertical wave-maker in the presence of surface tension}, 
volume={70}, 
DOI={10.1017/S0305004100049926}, 
number={2}, 
journal={Mathematical Proceedings of the Cambridge Philosophical Society}, 
author={Rhodes-Robinson, P. F.}, 
year={1971}, 
pages={323–337}
}

@article{bush2006walking,
  title={Walking on water: biolocomotion at the interface},
  author={Bush, John W. M. and Hu, David L},
  journal={Annual Review of Fluid Mechanics},
  volume={38},
  pages={339--369},
  year={2006},
  publisher={Annual Reviews}
}

@article{roh2019honeybees,
  title={Honeybees use their wings for water surface locomotion},
  author={Roh, Chris and Gharib, Morteza},
  journal={Proceedings of the National Academy of Sciences},
  volume={116},
  number={49},
  pages={24446--24451},
  year={2019},
  publisher={National Academy of Sciences}
}

@article{ho2023capillary,
  title={Capillary surfers: Wave-driven particles at a vibrating fluid interface},
  author={Ho, Ian and Pucci, Giuseppe and Oza, Anand U and Harris, Daniel M},
  journal={Physical Review Fluids},
  volume={8},
  number={11},
  pages={L112001},
  year={2023},
  publisher={APS}
}

@article{longuet1977mean,
  title={The mean forces exerted by waves on floating or submerged bodies with applications to sand bars and wave power machines},
  author={Longuet-Higgins, Michael Selwyn},
  journal={Proceedings of the Royal Society of London. A. Mathematical and Physical Sciences},
  volume={352},
  number={1671},
  pages={463--480},
  year={1977},
  publisher={The Royal Society London}
}

@book{glycerine1963physical,
  title={Physical properties of glycerine and its solutions},
  author={{Glycerine Producers' Association}},
  year={1963},
  publisher={Glycerine Producers' Association}
}

@article{cheng2008formula,
  title={Formula for the viscosity of a glycerol-water mixture},
  author={Cheng, Nian-Sheng},
  journal={Industrial \& engineering chemistry research},
  volume={47},
  number={9},
  pages={3285--3288},
  year={2008},
  publisher={ACS Publications}
}

@article{yeung1981added,
  title={Added mass and damping of a vertical cylinder in finite-depth waters},
  author={Yeung, Ronald W},
  journal={Applied Ocean Research},
  volume={3},
  number={3},
  pages={119--133},
  year={1981},
  publisher={Elsevier}
}

\end{document}